\documentclass[aps,prd,twocolumn,superscriptaddress,nofootinbib,10pt,reprint,preprintnumbers]{revtex4-1}

\usepackage[utf8]{inputenc}
\usepackage{graphicx,amsmath,amsfonts,amssymb,aas_macros,slashed}
\usepackage{bbold}
\usepackage{datetime}
\usepackage{numprint}
\usepackage{enumitem}
\usepackage{booktabs}
\usepackage{numprint}

\usepackage{graphicx,color,amsmath}
\usepackage{hyperref}

\allowdisplaybreaks

\newcommand{\MeV}{\ensuremath{\mathrm{MeV}}}
\newcommand{\GeV}{\ensuremath{\mathrm{GeV}}}

\newcommand{\ttt}{\ensuremath {3\to 2}}
\newcommand{\ftt}{\ensuremath {4\to 2}}

\DeclareMathOperator{\Ei}{Ei}

\usepackage{pgfplots}
\usepackage{bm}
\usepackage{mathtools}

\begin{document}
\preprint{UWThPh 2024-5}

\title{Even SIMP miracles are possible}

\author{Xiaoyong Chu}
\email{chuxiaoyong@ucas.ac.cn}
\affiliation{Institute of High Energy Physics, Austrian Academy of Sciences, Dominikanerbastei 16, 1010 Vienna, Austria}
\affiliation{International Centre for Theoretical Physics Asia-Pacific, University of Chinese Academy of Sciences, 100190 Beijing, China}

\author{Marco Nikolic}
\affiliation{Institute of High Energy Physics, Austrian Academy of Sciences, Dominikanerbastei 16,  1010 Vienna, Austria}
\affiliation{University of Vienna, Faculty of Physics, Boltzmanngasse 5, A-1090 Vienna, Austria}

\author{Josef Pradler}
\email{josef.pradler@oeaw.ac.at}
\affiliation{Institute of High Energy Physics, Austrian Academy of Sciences, Dominikanerbastei 16,  1010 Vienna, Austria}
\affiliation{University of Vienna, Faculty of Physics, Boltzmanngasse 5, A-1090 Vienna, Austria}

\begin{abstract}
Strongly interacting massive particles $\pi$ have been advocated as prominent dark matter candidates when they regulate their relic abundance through odd-numbered $3 \pi \to2\pi$ annihilation. 
We show that successful freeze-out may also be achieved  through {\it even-numbered} interactions $X X \to \pi \pi $  once  bound states $X$ among the particles of the low-energy spectrum exist. In addition, $X$-formation hosts the potential of also {\it catalyzing} odd-numbered $3 \pi \to2\pi$ annihilation processes, turning them into effective two-body processes $\pi X \to \pi\pi$. 
Bound states are often a natural consequence of strongly interacting theories. We calculate the dark matter freeze-out and comment on the cosmic viability and possible extensions. Candidate theories can encompass confining sectors without a mass gap, glueball dark matter, or $\phi^3$ and $\phi^4$ theories with strong Yukawa or self-interactions.

\end{abstract}

\maketitle

\paragraph*{Introduction.}

The several decades-long efforts to  detect dark matter (DM)  non-gravitationally have, to a significant degree, been fuelled by a relic density argument: the couplings of DM to the Standard Model (SM) that allow for phenomenological exploration also successfully generate the DM abundance in the early Universe through thermal two-body annihilation of DM into SM states.
As has been realized more recently, when the link between DM and SM becomes too weak, DM may still regulate its abundance through \ttt\ DM-only processes~\cite{Hochberg:2014dra}; also~\cite{Dolgov:1980uu, Carlson:1992fn, deLaix:1995vi}. This offers a pathway of DM as a thermal relic even when being partially secluded from~SM. 
The odd-numbered \ttt\ reaction is  naturally realized through the Wess-Zumino-Witten (WZW) five-point interaction of a strongly interacting dark sector~\cite{Hochberg:2014kqa}. 
The possibility of a DM number-depleting process that proceeds without participation of additional degrees of freedom is hence very attractive and has led to a flurry of further exploration, see~\cite{Hochberg:2015vrg, Kuflik:2015isi,Bernal:2015bla, Bernal:2015xba, Bernal:2015ova, Kuflik:2015isi, Choi:2015bya, Choi:2016hid,Soni:2016gzf, Kamada:2016ois,  Bernal:2017mqb,Cline:2017tka, Choi:2017mkk,  Kuflik:2017iqs, Heikinheimo:2018esa, Choi:2018iit,Hochberg:2018rjs, Bernal:2019uqr, Choi:2019zeb,  Katz:2020ywn, Smirnov:2020zwf, Xing:2021pkb, Braat:2023fhn, Bernreuther:2023kcg} among others.

{In this letter, we show that  bound states $X\equiv [\pi\pi]$ among strongly interacting massive particles (SIMPs), denoted as $\pi$, impact relic abundance predictions, potentially altering the conventional understanding of the ``SIMP mechanism''. In particular, $X$ may lead to a catalysis of freeze-out reactions by adding new channels}
\begin{subequations}
\begin{align}
\label{eq:cat-wzw} \!\!\text{\textit{catalyzed} \ttt\ annihilation: }  & \    \pi + X \to \pi +  \pi \,, 
 \\
\label{eq:cat-ftt}
\!\!\text{\textit{catalyzed} \ftt\ annihilation: } & \  X +  X\to \pi + \pi\, \,.
\end{align}
\end{subequations}
The last two are effective $2\to 2$ processes and compete with the free $3\pi\to 2\pi $ and $4\pi\to 2\pi$ counterpart reactions in depleting the overall DM mass density. Moreover, whereas $3\pi\to 2\pi$ and \eqref{eq:cat-wzw} are related through the same underlying odd-numbered interaction, the final process $X  X\to \pi \pi $ can be entirely due to \emph{even-numbered} interactions, such as the four-point self-interaction. This releases a requirement on the interaction structure of the theory and opens the door to a SIMP mechanism
without relying on anomaly-mediated interactions.

\begin{figure}[t]
\includegraphics[width=\columnwidth]{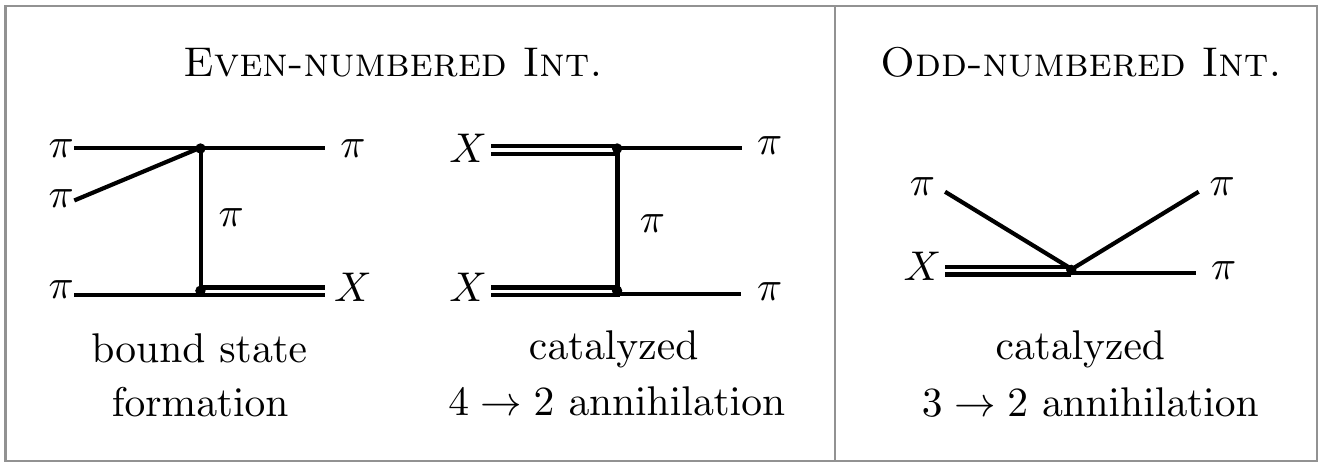}
\caption{
$X$-formation;
$X$-assisted annihilation via two 4-point interactions; catalyzed $3\to 2 $ annihilation.}
\label{fig:scheme}
\end{figure}

Of course, the prospect of catalyzed reactions~\eqref{eq:cat-wzw} and~\eqref{eq:cat-ftt} taking place requires~$X$ to be part of the low-energy spectrum. Once the theory allows for the existence of sufficiently long-lived~$X$, their formation is guaranteed through the radiationless exoergic process
\begin{align}
\label{eq:guaranteed-formation}
     \text{\textit{guaranteed} $X$ formation:} &\quad  \pi +  \pi +  \pi \to   \pi + X  \,.
\end{align}
This reaction may also be mediated through even-numbered interactions, and in its effective strength, it is not suppressed relative to the standard $3\to 2$ process. Hence, $X$ may form  efficiently, and it shows that already for models of SIMPs in isolation, the role of bound states calls to be studied; see Fig.~\ref{fig:scheme} for illustration.

\paragraph*{\boldmath Exemplary SIMP model.}
\label{sec:model}

To allow for a paralleling exposition close to the original papers on the SIMP-mechanism~\cite{Hochberg:2014dra,Hochberg:2014kqa} we shall consider the low-energy effective theory of $a= 1,\dots, N_\pi$ massive pseudo-Goldstone bosons $\pi_a$ as the DM candidates emerging from a confining dark non-Abelian gauge group of $N_f$ fermion fields.
The dynamics discussed is not exclusive to this choice and further possibilities will be commented on.

In the construction of the chiral Lagrangian,~$\pi_a$ are written as fluctuations  of the orientation of the chiral condensate $\Sigma_0$, $\Sigma= e^{i \pi/f_\pi} \Sigma_0 e^{i \pi^T/f_\pi}$, with $\pi = \sum_{n} \pi_n T^n$ where $T^n$ are the $N_\pi$ broken generators of the flavor group with normalization $\operatorname{Tr}[T^a T^b] = \delta^{ab}/2 $. 
Expanding in terms of $\Sigma$ yields the canonically normalized kinetic terms, masses, and  even-numbered interactions of~$\pi$. Considering a flavor-degenerate quark mass matrix $M$  with entries $m$, 
their universal mass is given by $m_\pi^2=\pm {2\mu^3 m}/{f_\pi^2}$; $f_\pi $ is the decay constant, and the plus (minus) sign applies to $Sp(2N_f)$ ($SU(N_f)$ or $SO(N_f)$) residual flavor symmetry.
Interactions are given by, 
\begin{align}
\label{eq:chiralLint}
    \mathcal{L}^{\rm even}_{\rm int} & \supset  -\frac{1}{3 f_{\pi}^{2}} \operatorname{Tr}\left(
[\pi,\partial_\mu \pi][\pi,\partial^\mu \pi]   
    \right)
    +\frac{m_{\pi}^{2}}{3 f_{\pi}^{2}} \operatorname{Tr}\left[ \pi^{4} \right] 
\end{align}
plus higher order terms $O(\pi^6/f_\pi^6)$.  Odd-numbered interactions in form of a non-vanishing WZW term are only present for symmetry-breaking pattern with coset spaces with non-trivial fifth homotopy groups~\cite{Witten:1983tw}. The leading order WZW Lagrangian then reads,\
\begin{align}
\label{eq:LWZW}
    \mathcal{L}_{\rm int}^\text{odd} & =\frac{2 N_{c}}{15 \pi^{2} f_{\pi}^{5}} \epsilon^{\mu \nu \rho \sigma} \operatorname{Tr}\left[\pi \partial_{\mu} \pi \partial_{\nu} \pi \partial_{\rho} \pi \partial_{\sigma} \pi\right].
\end{align}

 In the picture of strongly interacting theories, $X$ would be a  ``meson molecule'' or ``tetraquark'' of mass $m_X = 2m_\pi - E_B$ and $E_B > 0$. For the exposition of our ideas, we  assume a shallow bound molecule with $\kappa \equiv E_B/m_\pi \sim 0.1$ so that it can be treated as a non-relativistic bound state~\cite{Petraki:2015hla}. This points to a theory with $m/\Lambda \lesssim 1$ where $\Lambda \simeq 2\pi f_\pi$~\cite{Georgi:1992dw}. Theories with $m/\Lambda \ll 1$ such as for light quarks in QCD have a mass-gap whereas for $m/\Lambda \gg 1$, the lowest lying states are expected to be gluonia. The general ideas presented here also apply to deeper bound systems, but their treatment requires advanced field theoretical tools, greatly complicating matters. 
Even if we are far from the chiral limit, using~\eqref{eq:chiralLint} and~\eqref{eq:LWZW} allows for a most direct comparison with the original SIMP idea.
In the following, we shall take an $Sp(2N_c)$ gauge theory ($N_c=2$) for concreteness with $N_f=2$ fundamental Dirac fermions. After chiral symmetry breaking, the vacuum alignment is $\Sigma_0 = E $ where $E$ is the unitized invariant matrix of the remaining $Sp(4)$ flavor symmetry group. A detailed study of this choice is presented in~\cite{Kulkarni:2022bvh}  {(see also previous lattice studies~\cite{Bennett:2017kga,Bennett:2019jzz}) and there are works studying the existence of~$X$ in such scenarios, {\emph e.g.}~\cite{Zouzou:1986qh, Heupel:2012ua, Czarnecki:2017vco, DarisInprep}.} 
Another strongly-interacting option is taking $m/\Lambda \gg1$ with glueball~DM~\cite{Faraggi:2000pv,Soni:2016gzf, Yamanaka:2019aeq, Yamanaka:2019yek} and to consider their bound states~\cite{Giacosa:2021brl}. This is left for future work~\cite{upcoming}.

\paragraph*{Bound state-assisted SIMP annihilation.}
Before a detailed analysis, simple  estimates may convince us that  bound state formation  and $X$-assisted annihilation are both efficient and may even supersede odd-numbered interactions. 
The parametric ratio of rates of $X$ formation to odd-numbered annihilation reads,
\begin{align}
\label{eq:XformationvsWZW}
    \frac{\Gamma_{3\pi\to  \pi X}}{\Gamma_{3\pi\to 2\pi}}  & = \frac{\langle \sigma_{3\pi\to  \pi X} v^2 \rangle  }{\langle\sigma_{3\pi\to 2\pi} v^2 \rangle } \approx \frac{|\psi(0)|^2 f_\pi^2 } {m_\pi^5}   x_{\rm f}^2 .
\end{align}
Here $\langle \sigma_{i} v^2 \rangle $ are the thermally averaged collision integrals of the respective processes.
The dimensionful factor that relates both ``cross sections'' is the square of the bound state wave function at the origin $|\psi(0)|^2$; $x^2_{\rm f} \equiv (m_\pi/T_{\rm f})^2 \sim 400$ is an enhancement factor that accounts for the different velocity scalings of rates,  {$D$-wave for $3\pi\to 2\pi$ and $S$-wave for $3\pi \to \pi X$},  at non-relativistic freeze-out temperature $T_{\rm f} \sim m_\pi/20$. Taking as an estimate $|\psi(0)|^2 = 1/a_{B}^3$ with the Bohr radius $a_{B}=2/(\alpha_s' m_\pi)$ given in terms of the dark strong coupling constant $\alpha_s' = O(1)$ shows that the ratio in~\eqref{eq:XformationvsWZW} easily exceeds unity on account of $m_\pi/f_\pi = O(1)$ in strongly interacting theories\,\footnote{Alternatively, considering a spherical well $V(r)= - V_0 \theta (r -r_0)$, setting $V_0 \simeq  3.16 m_\pi$  and $r_0 = m_\pi^{-1}$, results in a binding energy $E_B = 0.1 m_\pi$ and yields $R(0)=\sqrt{4\pi} |\psi(0)| \simeq   m_\pi^{3/2}$~\cite{Mahbubani:2020knq}. }.

We may also convince ourselves that $X$-assisted annihilation competes with its free counterpart. Naive dimensional analysis suggests that, for a pair of $\pi$-particles, it is more likely for them to meet as constituents of a bound state than as free particles,
\begin{align}
\label{eq:catalysis}
   \frac{n_X |\psi(0)|^2  }{n_\pi^2} \approx 2\sqrt{2}\pi^{3/2} x_{\rm f}^{3/2} e^{\kappa x_{\rm f}}  \frac{|\psi(0)|^2}{m_\pi^3}\, ,
\end{align}
where we used non-relativistic Maxwell-Boltzmann statistics for~$n_\pi$ and $n_X$. 
Note that $2\sqrt{2}\pi^{3/2} x_{\rm f}^{3/2} \approx 10^3$ for $x_{\rm f}=20$.  Thus, the ratio in~\eqref{eq:catalysis} may easily exceed unity and suggests on general grounds that ${X X \to \pi\pi}$ dominates over ${4\pi\to 2\pi}$, and that ${\pi X \to \pi\pi}$ dominates over ${3\pi\to 2\pi}$ when odd-numbered interactions are present. This is what we mean by ``catalysis''.

\paragraph*{Cross sections.}
\label{sec:cs}
We now calculate the relevant cross sections from~\eqref{eq:chiralLint} and~\eqref{eq:LWZW} and first consider the $X$-formation cross section $(\sigma_{3\pi\to \pi X}v^2)$\,\footnote{See the Supplemental Material for detailed calculation of interaction rates, which
includes Refs.~\cite{Bhattacharya:2019mmy, Hochberg:2014kqa, Kamada:2022zwb, Petraki:2015hla, Peskin:1995ev}.}. In the approximation that we are working in, the amplitude for a bound state process is obtained from the free amplitude as follows. {Following the notation of \cite{Peskin:1995ev} and calling} $\langle \vec k_1, \vec k_2, \vec k_3  | \mathcal T \{ \dots \} | \vec p_1, \vec p_2, \vec p_3 \rangle$ the matrix-element of $3\pi \to 3\pi $ with respective incoming and outgoing momenta $\vec p_i$ and $\vec k_i$, the amplitude for the bound state formation $3\pi \to \pi X$ is obtained from 
\begin{align}
\label{eq:BSamplitude}
 \frac{\sqrt{2M_X}}{2m_\pi}\!\! \int \!\!\frac{d^3\vec q}{(2\pi)^3} \Big\langle \frac{\vec K}{2} - \vec q, \frac{\vec K}{2} +\vec q, \vec k_3  \Big| \mathcal T \{ \dots \} \Big| \vec p_1, \vec p_2, \vec p_3 \Big\rangle \tilde \psi(\vec q).
\end{align}
Here, $\vec K$ is the three-momentum of $X$; $\tilde \psi(\vec q)$ is the Fourier transform $ \tilde \psi(\vec q) = \int d^3\vec x\ \psi(\vec x) e^{-i \vec q\cdot \vec x} $ of the wave function that is a solution to the non-relativistic Schr\"odinger equation with confining potential $V(r)$ where $r = |\vec x|$ is the separation of the constituent SIMPs.

On general grounds, many terms contribute to the integral in~\eqref{eq:BSamplitude}. They can be classified by the power of the Cartesian components of $q_i$ that enter through the matrix element, enforcing a selection rule on $\psi(\vec x)$. Decomposing the latter into  angular and radial parts, $\psi(\vec x) = Y_{\ell m}(\hat x)R(r)$, constant terms $(q_i)^0$ yield for the integral $\psi(\vec x=0)$ which is only non-vanishing for $S$-states $\ell = 0$ where $R(0) =\sqrt{4\pi}|\psi(0)| \neq 0$. This is our most important case, as it concerns the ground state of~$X$. Terms proportional to $q_i$ yield the derivative of the radial wave function at the origin, $R'(0)\equiv dR/dr|_{r=0}$, so the lowest contributing angular momentum state is $P$-wave with $\ell =1$. We will encounter this case for the WZW interaction below.  Finally, terms quadratic in $q_i$ yield nominally divergent integrals in~\eqref{eq:BSamplitude}, probing the short distance behavior of the theory. In dimensional regularization one may show that $\nabla^2 R(0)=-m_\pi E_B R(0)$ holds~\cite{Brambilla:2002nu,Biondini:2021ycj}. 
Since we take $ E_B \ll m_\pi$, this becomes subleading in processes involving $S$-states, and we are hence allowed to neglect such contributions.

For the bound state formation process $3\pi \to \pi X $, there are then various diagrams to consider. During non-relativistic freeze-out, the dominant processes are $t$- and $u$-channel type diagrams of the sort depicted in Fig.~\ref{fig:scheme} where two 4-point interactions are connected via a SIMP propagator. The denominators of the propagators  are enhanced by a matching kinematic condition $k^2-m_\pi^2 \propto - E_B m_\pi $.
This renders other diagrams
irrelevant. 
The cross section is then given by 
\begin{equation}
\label{eq:sigmaBSF}
 \langle\sigma_{3\pi\to \pi X}v^2\rangle \simeq 
 \frac{\numprint{57041}}{\numprint{1310720} \sqrt{3}\pi^2} \frac{R_S^2(0)}{ f_\pi^8} \left( \frac{m_\pi}{E_B} \right)^{3/2} \, .
\end{equation}
The prefactor depends on the gauge group and symmetry breaking pattern and is averaged all possible incoming and summed over all outgoing flavor-combinations so that $n_\pi^3\langle\sigma_{3\pi\to \pi X}v^2\rangle $ yields the total number change per time.
In obtaining the result, we used a non-relativistic expansion, assuming that the typical incoming kinetic energy satisfies  \( m_\pi \langle v^2\rangle/2 \lesssim E_B \). This is equivalent to demanding  $T_{\rm f}\lesssim E_B$. The thermal average over a Maxwell-Boltzmann ensemble was taken in the final step. When comparing with~\eqref{eq:XformationvsWZW} we observe an additional enhancement by a factor of $( {m_\pi}/{E_B} )^{3/2}$ in the ratio of rates relative to the WZW-mediated $3\pi \to 2\pi$ annihilation. Detailed calculations of all cross sections are provided in the supplementary material to this letter.

Similarly, we may proceed with the calculation of the annihilation cross section for $XX\to \pi \pi$. A general computation would be a formidable challenge, but we may again profit from the imposed selection rules, focusing on $S$-wave initial bound states. There are six $t$- and $u$-channel diagrams, which become related in the limit that we neglect the internal motion of the constituents. The final  result reads
\begin{align}
\label{eq:XXtopipi}
\langle\sigma_{XX\to \pi\pi} v \rangle \simeq \frac{\numprint{2529757}}{\numprint{424673280}\sqrt{3}  \,\pi^3}\frac{R_S^4(0) }{ f_\pi^8 } \,,  %
\end{align}
where we have again summed over all flavor combinations.
Finally, when considering odd-numbered 
$3\to 2 $ interactions enabled by~\eqref{eq:LWZW},  we obtain the cross section for the related $\pi X \to \pi\pi$ process as
\begin{align}
\label{eq:BoundWZW}
\langle \sigma_{\pi X \to \pi\pi} v \rangle \simeq \frac{\sqrt{5} N_c^2 m_\pi^3}{512 \pi^6 f_\pi^{10} x} R_P'^2(0)  \, .
\end{align}
Importantly, the process requires $X$ to be in a $P$-wave state $X_P$ with $\ell =1$.
\begin{figure}[t]
\includegraphics[width=\columnwidth]{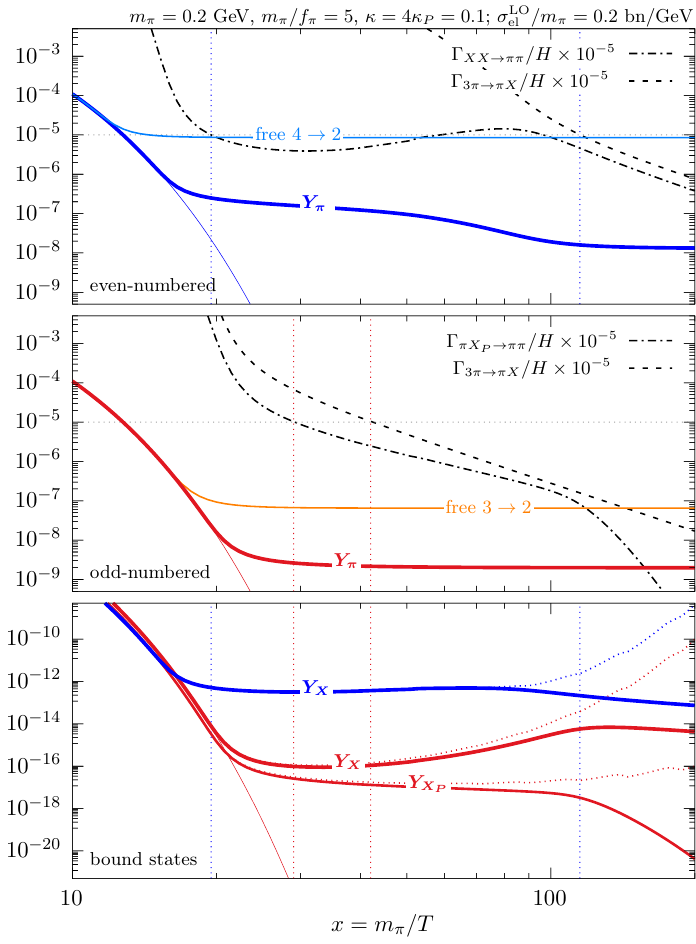}
\caption{Evolution of the DM abundance $Y_{\pi} + 2Y_{X}\simeq Y_{\pi} $
for $m_\pi = 0.2\,$GeV,   $m_\pi/f_\pi = 5$, $\kappa_S= 4\kappa_P =0.1$, $R(0)=0.3m_\pi^{3/2}$, and $dR(0)/dr=0.06m_\pi^{5/2}$. Thin solid and dotted lines show the  equilibrium  and detailed balance  abundances.
{\it Top:} even-numbered case where $XX$ annihilation freezes out at $x_1 \simeq 20$ (left vertical line) and bound-state formation freezes out at $x_2 \simeq 115$ (right vertical line); corresponding rates normalized to $H$ are shown as labeled. 
{\it Middle:} odd-numbered WZW interactions included. $X_P$ annihilation freezes out comparatively later (left vertical line), maintaining  longer chemical equilibrium. 
{Free $4\to 2$ and $3\to2$ freeze-out are  shown for comparison in the top and middle panel.}  
{\it Bottom:} associated bound state abundances with matching colors.
For the choice of parameters, the  DM abundance is reached for the odd case; using instead $R(0)=1.4m_\pi^{3/2}$ yields the DM abundance for the even case.}
\label{fig:sol}
\end{figure}

\paragraph*{Abundance evolution.}
\label{sec:sol}
We are now in a position to solve the evolution equations for the two populations, free $\pi$ and $X$. Their respective total comoving number densities, normalized to the total entropy density~$s$, are given by $ Y_{\pi,X} = n_{\pi,X}/s$, where $s$ is the total entropy density of the Universe and a sum over all flavors is implicit.
We assume that kinetic equilibrium with SM is maintained; we develop on this in the next section.
At high temperatures (small $x$), $\pi$ and $X$ follow their equilibrium distributions, $Y_{\pi,X} = Y_{\pi,X}^{\rm eq}$, due to fast number-changing processes. Their chemical decoupling happens at $x_{1} \sim 20$ when 
$n_X^2  \langle \sigma_{XX\to \pi \pi} v\rangle /n_\pi \simeq H(x_{1})$. Subsequently, considering only even-numbered interactions, assuming dominance of $XX\to \pi \pi$ over the free $4\pi \to 2\pi $ counterpart and neglecting the inverse process, a particularly simple form of the Boltzmann equation is found for the combination $Y_\pi + 2 Y_X$,%
\footnote{The approximate evolution of the overall mass density in the dark sector is $\rho_{\rm DM} \simeq  (Y_\pi + 2 Y_X)m_\pi s $ up to corrections $O(E_B/m_\pi)$.}
\begin{align}
\label{eq:Boltzsum}
  \frac{d(Y_\pi + 2 Y_X)}{dx} = - \frac{2  \langle \sigma_{XX\to \pi\pi} v \rangle Y_X^2  s(x)}{x H(x)} \quad (x > x_{1}) \, .
\end{align}
The immediate evolution that ensues for $x>x_{1}$ is non-trivial because bound state formation $3\pi \leftrightarrow \pi X$ is still operative. This maintains a detailed balance between the $X$ and $\pi$ populations,
\begin{equation}
\label{eq:detailed}
    Y_X  = \frac{Y_\pi^2 Y_X^{\rm eq}}{(Y_\pi^{\rm eq})^2} = Y_\pi^2 \frac{N_X}{N_\pi^2} (2\pi x)^{3/2} e^{\kappa x} \left( \frac{m_X}{m_\pi} \right)^{3/2} \frac{s(x)}{m_\pi^3}  \,,
\end{equation}
where $N_\pi =5$ and  $N_X = (N_\pi+1)N_\pi/2 = 15$ are the possible flavor combinations. Together,~\eqref{eq:detailed} and~\eqref{eq:Boltzsum} determine the evolution of $Y_\pi$ and $Y_X$ for as long as their detailed balance holds until bound state formation freezes out at $x_2>x_1$ when $\Gamma_{3\pi \to \pi X} \equiv n_\pi^2 \langle \sigma_{3\pi \to \pi X} v^2 \rangle  = H(x_2)$.

Using~\eqref{eq:detailed} in~\eqref{eq:Boltzsum} with $Y_X\ll Y_\pi$ and neglecting $dY_X/dx$, the Boltzmann equation becomes one for $dY_\pi/dx$ that can be integrated. To leading order in $x_1/x_2$ we obtain the following solution, 
\begin{align}
 & Y_\pi^{-3}(x_2) \simeq   \frac{256 \sqrt{2} \pi^8 g_*^{5/2} m_\pi M_P   \langle \sigma_{XX\to \pi\pi} v \rangle}{6075\sqrt{5} x_2^4 }  \frac{N_X^2}{N_\pi^4}  \notag\\
  & \times \left[  8(\kappa x_2)^4  \Ei(2\kappa x_2) - e^{2\kappa x_2} ( 3+2\kappa x_2 + 2 \kappa^2 x_2^2 +4\kappa^3 x_2^3 )  \right],
 \label{eq:evenBoltzSol}
\end{align}
where $\Ei$ is the exponential integral function and $M_P$ is the reduced Planck mass and $g_*$ are the effective degrees of freedom at~$x_2$.
This approximation works for $\kappa x_1\gtrsim 1 $ which is congruent with assuming $E_B >T$ at chemical decoupling. In writing the solution, we have also taken $Y^{-3}_\pi(x_1) \ll Y^{-3}_\pi(x_2) $. To within a factor of two, we may then put the solution in suggestive form,
\begin{align}\label{eq:BoltzSolApp}
\Omega^{\rm even}_\pi \sim  0.2
\left( {\frac{200\,\kappa x_2^5}{e^{2\kappa x_2}}} 
\frac{{\rm bn/GeV}}{ \langle \sigma_{XX\to \pi\pi} v \rangle/m_\pi} \frac{m_\pi}{\text{GeV}}\right)^{1/3} . 
\end{align}
This is a central result. First, note that  the relic density depends on $x_2$, i.e., the moment of freeze-out of bound state formation, and not~$x_1$. Second, we observe a strong dependence on $\kappa$, and with $x_2$ typically between 50 to~100, $\kappa =0.1$ suggests sub-GeV DM with cm$^2$/gram self-interactions---in the same ballpark as the odd-numbered SIMP case. Finally, when compared to ordinary $2\to 2$ freeze-out, the inverse dependence on the annihilation cross section is softened by the cubic root.

\begin{figure}[t]
\includegraphics[width=\columnwidth]{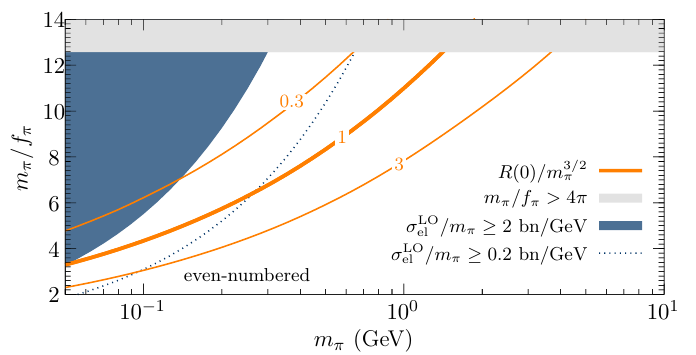}
\caption{{Contours of observed DM abundance in the even-numbered case for  $\kappa =0.1$ and $R(0)/m_\pi^{3/2}= 0.3,\,1,\,3$, respectively. The colored region (dotted line) shows the DM self-scattering constraint $\sigma^{\rm LO}_{\rm el} \ge 2$\,($0.2$)\,bn/GeV. The free $4\pi \to 2\pi$ counterpart would require $m_\pi/f_\pi > 4\pi$ for $m_\pi > 1$\,MeV.}
}
\label{fig:evenscan}
\end{figure}

The top panel of Fig.~\ref{fig:sol} shows the full numerical solution for $Y_\pi$ and $Y_X$ as a function of~$x$ for the parameters given in the caption.
The evolution contains three steps as discussed above. First, chemical decoupling happens at $x_1\simeq 20$ when $XX \leftrightarrow \pi\pi$ freezes out. Both $X$ and $\pi$ develop a chemical potential and keep a detailed balance  via $3\pi \leftrightarrow \pi X$ until $x_2\simeq 115$. For $x>x_2$ free $\pi$ have reached their relic density value while $Y_X$ keeps decreasing further, as the rate of $X X \to \pi \pi$ with respect to $n_X$ is still larger than the Hubble rate, $n_X \langle \sigma_{XX\to \pi\pi} v \rangle  > H(x_2)$. The evolution of $Y_X$ is shown in the bottom panel. {Figure~\ref{fig:evenscan} shows that with bound states, even-numbered interactions can achieve the observed DM abundance with bn/GeV-scale self-scattering and reach $m_\pi>{\rm GeV}$ with perturbative couplings~\footnote{A comparison with the free $4\pi\to 2\pi$ and  $3\pi\to 2\pi$ freeze-out, as well as the parameter scan varying $\kappa$ and for the odd-numbered case, are given in the supplementary material.}.
}

\paragraph*{Odd-numbered case.}

We now turn our attention to the scenario when we are additionally afforded odd-numbered interactions. 
As calculated in~\eqref{eq:BoundWZW}, the efficiency of $\pi X_P\to \pi\pi$ entirely hinges on the availability~$X_P$ which must be present in the low energy spectrum. Since the path to collisional excitation is open, one may consider the detailed balancing relation  $n_{X_P}/n_{X_S}=3\exp[-|\kappa-\kappa_P|x] $ as an estimate for the number density $n_{X_P}$ of excited states;  $\kappa_P=E_P/m_\pi$ with $E_P$ being the $P$-wave binding energy.
As we shall see now, the impact of bound states~$X$ can also be substantial. Even with the bottleneck of $P$-wave states for WZW interactions, $\pi X_P\to \pi\pi$ supersede the free $3\pi \to 2\pi$ scenario, and is generally stronger than~$XX\rightarrow \pi \pi$. 

For the odd-numbered case, the chemical decoupling happens when $n_{X_P}  \langle\sigma_{\pi X_P \rightarrow \pi\pi} v\rangle  \simeq H(x_1) $. For $x \ge x_1$, the right hand side
 of~\eqref{eq:Boltzsum} is replaced by $- s 
\langle\sigma_{\pi X_P \rightarrow \pi\pi} v\rangle  Y_{\pi}Y_{X_P}/(x H)$. Since $Y_{X_P}/Y_{X_S}$, and thus the rate of $\pi X_P\to \pi\pi $, decrease exponentially for $x>x_1$,  $Y_\pi$ already freezes out at~$x_1$. If $ 3\pi \to \pi X$ decouples later, $x_2 > x_1$, \eqref{eq:detailed} allows us to estimate the yield from 
$  n_{X_P}^{\rm eq} \langle\sigma_{\pi X_P \rightarrow \pi\pi} v\rangle /H(x_1) \simeq  (Y_\pi^{\rm eq})^2/Y^2_\pi(x_1) 
$. 
This gives %
\begin{equation}
\label{eq:oddBoltzSol}
   \Omega^{\rm odd}_\pi \simeq  0.2 \,\left(\frac{ x_1}{20}\right)^{5/4} \left(\frac{e^{-\kappa_P x_1}\,10^{-3} \,{\rm bn/GeV}}{\langle\sigma_{\pi X_P \rightarrow \pi\pi} v\rangle/m_\pi} \right)^{1/2}\,. 
\end{equation}

The middle panel of Fig.~\ref{fig:sol} shows the full numerical freeze-out solution. In comparison with the even-numbered case, the stronger $\pi  X_P \to \pi\pi$ reaction maintains longer chemical equilibrium. At the same time, with $p$-wave states diminishing more rapidly with temperature, there is no distinct intermediate phase, and freeze-out happens in one step (unless considering large values for $dR(0)/dr$). Also shown is the standard SIMP scenario through free $3\to 2$ reactions, and we observe an order of magnitude smaller freeze-out yield for the chosen set of parameters when $X$ is considered (catalysis). This softens the notorious tension between maximum permissible elastic scattering cross section and relic density requirement in SIMP models~\cite{Hansen:2015yaa}, and for the shown case, both requirements are indeed satisfied. The leading order elastic scattering cross section is $\sigma_{\rm el}^{\rm LO}/m_\pi = 0.2$~bn/GeV, receiving higher order corrections in the chiral expansion~\cite{Hansen:2015yaa} as well as from the bound state in the spectrum. For the latter, we may estimate a $S$-wave scattering length of the order $1/(\kappa_S m_\pi^2)^{1/2}$~\cite{Braaten:2013tza}, leading to resonant-induced $\sigma_{\rm el}/m_\pi  \sim 1/(\kappa_S m_\pi^3)$. For $m_\pi =0.2\,$GeV and $\kappa_S \sim 0.1 $ adopted here, it suggests an elastic scattering in the same ballpark and below~bn/GeV.

Within the exemplary scenario of pseudo-Goldstone bosons making~$X$, we may also comment on the influence of additional low-lying states, such as $\rho$-mesons of mass $m_\rho \lesssim 2m_\pi$. Additional annihilation channels become available, such as $3\pi \to \pi \rho^*    \to \pi\pi$~\cite{Choi:2018iit} or $3\pi \to \pi \rho$~\cite{Bernreuther:2023kcg}. For the parameter region of interest and unless one considers a finely-tuned resonance region  $m_\rho \simeq 2m_\pi$, we find that these processes are generally subleading to the $X$-mediated ones and we are allowed to neglect them.

\paragraph*{Coupling to SM and longevity of~$X$.} 

As is pertinent to all SIMP scenarios that freeze out through self-depletion, kinetic equilibrium with radiation 
must be maintained to achieve a {\it cold} DM scenario. 
An elastic scattering process $\pi\,{\rm SM}_i \to \pi\,{\rm SM}_i$ with rate $ \Gamma_{\pi\,{\rm  SM}} = \langle \sigma_{\pi\,{\rm SM}} c \rangle  n_i  > H $, 
that brings SM and dark sector into kinetic equilibrium during freeze-out, generally also enables $\pi\pi \to {\rm SM}_i \overline{\rm SM}_i$ annihilation. The SIMP mechanism then requires $\Gamma_{\rm ann}  = n_\pi \langle  \sigma_{\rm ann} v \rangle  < H $,  where $\sigma_{\rm ann}$ is the cross section for $\pi\pi \to {\rm SM}_i \overline{\rm SM}_i$. On the account of $n_i/n_\pi \gg 1$, where $n_i$ is the number density of a relativistic SM species, both conditions are generically satisfied~\cite{Hochberg:2014dra}.
In the current context, interactions of $\pi$ with SM may also destabilize~$X$ through $ X=[\pi\pi] \to {\rm SM}_i \overline{\rm SM}_i$ and we must ensure that for the decay rate $\Gamma_X < H$ holds until after freeze-out. 

Assuming $\sigma_{\rm ann} v \simeq const.$ and
noting that $|\psi(0)|^2 v$ has units of particle flux,
we may estimate the induced decay width of $X$ as $\Gamma_X \sim  |\psi(0)|^2 (\sigma_{\rm ann} v) $ where $(\sigma_{\rm ann} v)$ is the $\pi\pi \to {\rm SM}_i \overline{\rm SM}_i$ annihilation cross section. 
The longevity requirement $\Gamma_X/ H < 1 $ thereby translates into an upper limit on the annihilation cross section,%
\begin{align} 
\label{eq:sigmaAnn}
\sigma_{\rm ann} v \lesssim 10^{-3} {\rm pb}\  x^{-2} \left( \frac{m_\pi}{100\ \MeV} \right)^2 \frac{\MeV^3}{|\psi(0)|^2}\, .
\end{align}
In the simplest cases, such as contact interactions through a heavy mediator, elastic and annihilation cross sections are additionally related and in the same ballpark, $\sigma_{\pi\,{\rm SM}} c \sim \sigma_{\rm ann}v$. We may then use the bound in~\eqref{eq:sigmaAnn} to estimate the implied ceiling on the elastic scattering rate,
\begin{align}
 1\lesssim   \frac{\Gamma_{\pi\,{\rm SM}}}{H} \lesssim 
    \frac{10^6}{x^3}
     \left(\frac{m_\pi}{100\ \MeV} \right)^3 \frac{\MeV^3}{|\psi(0)|^2} \,.
\end{align}
This can be easily satisfied at freeze-out for $|\psi(0)| < m_\pi^{3/2}$.  We hence conclude that it is possible to retain kinetic equilibrium while maintaining sufficient longevity of~$X$ and paired with sub-Hubble two-body annihilation. Therefore, the model-building requirements for coupling the dark sector to the SM are not escalated compared to the standard SIMP mechanism, and one may use the options already entertained in the original work~\cite{Hochberg:2014dra}.

Finally, additional $X$ formation and breakup reactions may open when introducing couplings to SM. It is important to note that the detailed balancing condition \eqref{eq:detailed} between $X$ and $\pi$---being a Saha equation---remains unaltered. If the new processes  dominate over $3\pi \leftrightarrow \pi X$,~\eqref{eq:detailed} retains its validity {\it longer}, $x_2$ will be larger, and the relic density smaller. Hence, if anything, the introduction of SM-interactions harbor the potential to make $X$-assisted freeze-out even more efficient, adding a level of richness, without jeopardizing the overall picture. 

\paragraph*{Conclusions.} 
Bound-state-assisted self-depletion offers a novel approach to DM relic density generation. It supports an even-numbered relic SIMP mechanism and enhances odd-numbered counterparts. Both are realized in strongly interacting theories.
A broader study of the many aspects mentioned in this work, as well as the exploration of other particle-physics realizations giving rise to $X$, such as glueballs or strong Yukawa forces, will be the subject of upcoming work~\cite{upcoming}.

\paragraph*{Acknowledgements.} 
JP thanks M.~Pospelov for a useful discussion on QCD-like theories. This work was supported by the Research Network Quantum Aspects of Spacetime (TURIS) and by the FWF Austrian Science Fund research teams grant STRONG-DM (FG1).
Funded/Co-funded by the European Union (ERC, NLO-DM, 101044443).

\bibliography{Refs}

\appendix

\onecolumngrid
\clearpage
\section{Boltzmann equations}
Here we collect the relevant Boltzmann equations for $\pi$ and $X$. The right-hand side thereby provides the definition of the thermally averaged cross sections $\langle \sigma_i v \rangle$ and collision terms $\langle \sigma_i v^2 \rangle $ that are computed in the subsequent section. Our yield variables are defined as flavor-summed quantities
$   Y_{\pi} \equiv \sum_a {n_{\pi_a}}/{s}$, $Y_{X} = \sum_{a\leq b} {n_{X_{ab}}}/{s}$ where the sum is over flavors $a,b,\dots = 1,\dots N_\pi$.  The evolution of $\pi$ and $X$ is then governed by 
\begin{align}
\label{eq:BoltzmannPiA}
\frac{d Y_\pi}{d x} =
\frac{1}{3 H(x)}\frac{ds}{dx} &
\left[ 
-2  \langle \sigma_{XX\to 2\pi} v\rangle \left(Y_X^2 - (\frac{Y_\pi}{Y^{\rm eq}_\pi} Y_X^{\rm eq}  )^2 \right)  
+ 2s\langle\sigma_{3\pi\to \pi X} v^2\rangle  \left(  Y_{\pi}^3 -Y_{\pi} (Y_{\pi}^{\rm eq})^2\frac{ Y_X}{Y_X^{\rm eq}} \right) \right. 
\notag \\ &  \left.
\hspace*{-1.5cm}
-  \langle\sigma_{\pi X \to 2\pi}v \rangle \left(  Y_{\pi}Y_X -\frac{ Y_{\pi}^2}{Y_{\pi}^{\rm eq}} Y_{X}^{\rm eq} \right)
+  s \langle\sigma_{3\pi\to 2\pi} v^2\rangle  \left(  Y_{\pi}^3 -Y_{\pi}^2 Y_{\pi}^{\rm eq} \right)
+  2 s^2 \langle\sigma_{4\pi\to 2\pi} v^3\rangle  \left(  Y_{\pi}^4 -Y_{\pi}^2 (Y_{\pi}^{\rm eq})^2 \right)
\right] ,\\
\frac{d Y_{X}}{d x} =
\frac{1}{3 H(x)}\frac{ds}{dx} &
\left[
 2\langle \sigma_{XX\to 2\pi} v\rangle \left(Y_X^2 - (\frac{Y_\pi}{Y^{\rm eq}_\pi} Y_X^{\rm eq})^2 \right)
- s \langle\sigma_{3\pi\to \pi X} v^2\rangle  \left(  Y_{\pi}^3 -Y_{\pi} (Y_{\pi}^{\rm eq})^2\frac{ Y_X}{Y_X^{\rm eq}} \right) 
\right.
\notag \\ & \left.
\qquad\qquad 
+\langle\sigma_{\pi X \rightarrow 2\pi} v\rangle \left(Y_{\pi}Y_{X}-\frac{Y_{\pi}^2}{Y_\pi^{\rm eq}}Y_{X}^{\rm eq}\right) 
\right]\ , \label{eq:BoltzmannX}
\end{align}
where the first (second) lines of each equation is due to by even (odd) numbered interactions. 
In our principal mass region of interest $m_\pi = O(100\,\MeV)$ freeze-out happens late enough that to reasonable approximation, we may neglect the variation of the relativistic degrees of freedom (including DM) and use $ds/dx = - 3 s/x$ when seeking analytical solutions; in our numerical treatment, we account for the full evolution.
The dark matter mass density is then obtained from  the mass-weighted sum of their freeze-out yields,
\begin{align}
    \Omega_{\rm dm} =\frac{s_0}{\rho_{\rm c}}  (m_X Y_{X,0} + m_\pi Y_{\pi,0}) \, ,
\end{align}
where the ``0'' superscript denotes the quantities today. In practice, $Y_{X,0} \ll Y_{\pi,0}$ or $Y_{X,0} $ is vanishing altogether because $X$ has decayed.

\section{Cross sections and collision integrals}

\subsection{Interaction terms and Feynman Rules}

The expansion of the leading-order chiral Lagrangian
 \begin{align}
\label{eq:chiralL}
    \mathcal{L}_{\chi}&= \frac{f_\pi^2}{4} \text{Tr} \left[ \partial_\mu \Sigma \partial^\mu \Sigma^\dagger\right]- \frac{ \mu^3}{2} \left(\operatorname{Tr} \left[ M \Sigma\right]+
    h.c.\right) 
\end{align}
to six-point interaction terms reads and dropping  mass and canonically normalized kinetic terms reads,
\begin{align}
    \mathcal{L}_{\rm int}^{\rm even} & =   -\frac{2}{3 f_{\pi}^{2}} \operatorname{Tr}\left[\pi^{2} \partial^{\mu} \pi \partial_{\mu} \pi-\pi \partial^{\mu} \pi \pi \partial_{\mu} \pi\right]  + \frac{m_{\pi}^{2}}{3 f_{\pi}^{2}} \operatorname{Tr}\left[ \pi^{4} \right] 
    \notag\\ & \quad  +\frac{16}{45 f_{\pi}^{4}} \operatorname{Tr}\left[\pi^{2}\left(\pi^{2} \partial^{\mu} \pi \partial_{\mu} \pi-\pi \partial^{\mu} \pi \pi \partial_{\mu} \pi\right)\right]
    -\frac{2m_\pi^2}{45 f_\pi^4}\operatorname{Tr}\left[\pi^{6}\right]   + \dots 
    \\&
    =  -\frac{1}{12 f_{\pi}^{2}} \sum_{kn}\left(\pi_k\pi_k \partial_\mu \pi_n \partial^\mu \pi_n -\pi_k \partial_\mu \pi_k  \pi_n \partial^\mu \pi_n \right)
    + \frac{m_{\pi}^{2}}{48 f_{\pi}^{2}} \left(\sum_{k}\pi_k^2\right)^2 
     \notag  \\& \quad 
     +\frac{1}{180 f_{\pi}^{4}} \sum_{jkn} \pi_j^2\left(\pi_k\pi_k \partial_\mu \pi_n \partial^\mu \pi_n -\pi_k \partial_\mu \pi_k  \pi_n \partial^\mu \pi_n \right) - \frac{m_{\pi}^{2}}{2880 f_{\pi}^{4}} \left(\sum_{k}\pi_k^2\right)^3
     + \dots\,.
\end{align}
The sums run over $i,j,\dots = 1,\dots, N_\pi$. For  $\pi \in SU(4)/Sp(4)$, i.e.~specializing to $N_f =2$, there are $N_\pi = 5$ pseudo-Nambu-Goldstone fields.

A technical aspect of computing the relevant cross sections is the combinatorics of fields only distinguished by their flavor index. We may obtain the relevant Feynman rules (up to a global factor~$i$) for the four- and six-point interactions directly by contracting the respective operators with the multiparticle state $  | p_{1, a_1} p_{2, a_2}\dots  p_{m, a_m}\dots \rangle $, which is to be understood as an ordered list of momenta: $m$ labels the particle by its position in the multiparticle state and $a_m$ denotes its flavor. Each momentum 
 is defined with its direction pointing towards the interaction vertex.
We start with the simplest term, the quartic interaction with constant coupling $m_\pi^2/3f_\pi^2$,
\begin{align}
 \frac{m_{\pi}^{2}}{3 f_{\pi}^{2}} \langle 0 | \operatorname{Tr}\left[ \pi^{4} \right]  | p_{1, a_1} p_{2, a_2} p_{3, a_3}p_{4, a_4}  \rangle 
 & = \frac{m_{\pi}^{2}}{3 f_{\pi}^{2}} \sum_{ijkl} \operatorname{Tr}\left[ T^i T^j T^k  T^l \right] \langle 0 |  \pi_i \pi_j \pi_k \pi_l    | p_{1, a_1} p_{2, a_2} p_{3, a_3}p_{4, a_4}  \rangle \notag\\
 & = \frac{m_{\pi}^{2}}{3 f_{\pi}^{2}} \sum_{ijkl} \operatorname{Tr}\left[ T^i T^j T^k  T^l \right] \, \sum_{ \sigma } \delta_{i,  \sigma(a_1)}\delta_{j, \sigma(a_2)}\delta_{k, \sigma(a_3)}\delta_{l, \sigma(a_4)}\,,\notag\\
  &  = \frac{m_{\pi}^{2}}{3 f_{\pi}^{2}} \sum_{ \sigma}  \operatorname{Tr}\left[ T^{\sigma(a_1)} T^{\sigma(a_2)} T^{\sigma(a_3)}  T^{\sigma(a_4)} \right]\,, 
\end{align}
where $\sigma = \sigma(a_1)\, \sigma(a_2)\,\sigma( a_3)\,\sigma( a_4)$ sums up all $4!$ permutations of $a_1\, a_2\, a_3\, a_4$. For the kinetic quartic interaction, derivatives $\partial^\mu \pi_{\sigma(a_m)}$ are replaced by $ip^\mu_{m, \sigma(a_m)} \pi_{\sigma(a_m)}$  and we write $p_{m, \sigma(a_m)} \equiv p_{m}$ below for brevity. We obtain,
\begin{align}
-\frac{2}{3 f_{\pi}^{2}}\langle 0 | \operatorname{Tr}\left[\pi^{2} \partial^{\mu} \pi \partial_{\mu} \pi-\pi \partial^{\mu} \pi \pi \partial_{\mu} \pi\right]   | p_{1, a_1} ... \rangle & = -\frac{2}{3 f_{\pi}^{2}}  \sum_{ijkl} \operatorname{Tr}\left[ T^i T^j T^k  T^l \right]  i^2 (p_k \cdot p_l - p_j \cdot  p_l)\langle 0 | \pi_i \pi_j \pi_k \pi_l    | p_{1, a_1} ... \rangle \notag\\
& = \frac{2}{3 f_{\pi}^{2}}   \sum_{ \sigma }  \operatorname{Tr}\left[ T^{\sigma(a_1)} T^{\sigma(a_2)} T^{\sigma(a_3)}  T^{\sigma(a_4)} \right] \,  (p_{3} \cdot p_{4} - p_{2} \cdot  p_{4}) \,,
\end{align}
For the concrete model of $SU(4)/Sp(4)$ the trace of generators of the coset space evaluates to
\begin{equation}
\operatorname{Tr}\left[ T^{a_1} T^{a_2} T^{a_3}  T^{a_4} \right] = {1\over 16 }\left( \delta_{a_1, a_2} \delta_{a_3, a_4}  +  \delta_{a_1, a_4} \delta_{a_2, a_3}  -  \delta_{a_1, a_3} \delta_{a_2, a_4}  \right)\,.
\end{equation}
This allows us to write the four-point vertex explicitly as
\begin{align}
\Gamma_{\left\{p_1, a_1\right\}, \left\{p_2, a_2\right\}, \left\{p_3, a_3\right\}, \left\{p_4, a_4\right\} } = \frac{1}{6f_\pi^2} \big\{ &\delta_{a_1, a_2}\delta_{a_3, a_4} \left( 2 p_1 \cdot p_2 +  2 p_3 \cdot p_4  -  p_1 \cdot p_3  -  p_1 \cdot p_4  -   p_2 \cdot p_3  - p_2 \cdot p_4 +m_\pi^2  \right)   \notag\\
   + &  \delta_{a_1, a_3}\delta_{a_2, a_4} \left( 2 p_1 \cdot p_3 +  2 p_2 \cdot p_4  -  p_1 \cdot p_2  -  p_1 \cdot p_4  -   p_2 \cdot p_3  - p_3 \cdot p_4 +m_\pi^2  \right) \notag\\
   +  &   \delta_{a_1, a_4}\delta_{a_2, a_3} \left( 2 p_1 \cdot p_4 +  2 p_2 \cdot p_3  -  p_1 \cdot p_2 -  p_1 \cdot p_3  -   p_2 \cdot p_4  - p_3 \cdot p_4  + m_\pi^2  \right) \big\}\,,
\end{align}
where the kinetic interaction induces the terms involving momentum scalar products, while the last terms proportional to $m_\pi^2$ are from ${\rm Tr}[\pi^4]$. 

Regarding Feynman rules of six-point interactions, we similarly obtain
\begin{equation}
 \sum_{\sigma} \left[ -\frac{2m_\pi^2}{45 f_\pi^4} - \frac{16}{45 f_\pi^4} (p_5\cdot p_6 - p_4 \cdot p_6 ) \right] \operatorname{Tr}\left[T^{\sigma(a_1)} T^{\sigma(a_2)} T^{\sigma(a_3)}  T^{\sigma(a_4)} T^{\sigma(a_5)}  T^{\sigma(a_6)}  \right]  \,,
\end{equation}
where $\sigma = \sigma(a_1)\, \sigma(a_2)\,\sigma( a_3)\,\sigma( a_4)\,\sigma( a_5)\,\sigma( a_6)$  sums up the $6!= 720$ permutations of $a_1\, a_2\, a_3\, a_4\, a_5\, a_6$. The prefactor 
is insensitive to switching the order of $a_1$ and $a_2$, i.e., for $SU(4)/Sp(4)$ we can use the equality
\begin{equation}
  \sum_{\sigma}  \operatorname{Tr}\left[{  T^{a_1} T^{a_2} + T^{a_2} T^{a_2}  \over 2 }  T^{a_3}  T^{a_4} T^{a_5}  T^{a_6}  \right] = \sum_{\sigma} {   \delta_{a_1, a_2} \over 8}    \operatorname{Tr}\left[ T^{a_3}  T^{a_4} T^{a_5}  T^{a_6}  \right]
\end{equation}
to reduce the number of generators in the trace. 
Then, replacing $a_i$ $(i=1,..,6)$ by $(a,b,c,d,f,g)$ results in 
\begin{align}
\Gamma_{\rm 6-point}|_{\rm contact} =  -  \frac{ m_\pi^2}{60 f_\pi^4}  &  ( \delta_{a,g} \delta_{b,f} \delta_{c,d}+ \delta_{a,f} \delta_{b,g} \delta_{c,d}  +  \delta _{a,g} \delta _{b,d} \delta _{c,f}+  \delta _{a,d} \delta _{b,g} \delta _{c,f}+ \delta _{a,f} \delta _{b,d} \delta _{c,g} + \delta _{a,d} \delta _{b,f} \delta _{c,g}\notag\\ 
&  +  \delta _{a,g} \delta _{b,c} \delta_{d,f}+ \delta _{a,c} \delta _{b,g} \delta _{d,f}+ \delta _{a,b} \delta_{c,g} \delta _{d,f}+ \delta _{a,f} \delta _{b,c} \delta _{d,g}  +  \delta _{a,c} \delta _{b,f} \delta _{d,g}+ \delta _{a,b} \delta _{c,f} \delta _{d,g}  \notag\\ 
& +  \delta_{a,b} \delta_{c,d} \delta _{f,g} +  \delta _{a,d} \delta _{b,c} \delta _{f,g}  + \delta _{a,c} \delta _{b,d} \delta _{f,g} ) \, 
\end{align}
for the contact interactions originating from the mass term in the chiral Lagrangian. For the six-point interaction originating from the kinetic term,  we further simplify the expression by tailoring it to the process $X_{ab} + X_{cd} \to \pi_f + \pi_g$, taking the non-relativistic limit for initial states (and thus $p_f \cdot p_g = 2m_X^2- m_\pi^2$ for final states) 
\begin{align}
\Gamma_{\rm 6-point}|_{\rm kinetic} =   \frac{ 13m_X^2 - 4 m_\pi^2}{180  f_\pi^4} &  ( \delta_{a,g} \delta_{b,f} \delta_{c,d}+ \delta_{a,f} \delta_{b,g} \delta_{c,d}
 +  \delta _{a,g} \delta _{b,d} \delta _{c,f}+  \delta _{a,d} \delta _{b,g} \delta _{c,f}+ \delta _{a,f} \delta _{b,d} \delta _{c,g} + \delta _{a,d} \delta _{b,f} \delta _{c,g}  \notag\\ 
&  +  \delta _{a,g} \delta _{b,c} \delta _{d,f}+ \delta _{a,c} \delta _{b,g} \delta _{d,f}+ \delta _{a,b} \delta _{c,g} \delta _{d,f}+ \delta _{a,f} \delta _{b,c} \delta _{d,g}  +  \delta _{a,c} \delta _{b,f} \delta _{d,g}+ \delta _{a,b} \delta _{c,f} \delta _{d,g}  \notag\\ 
& -4 \delta_{a,b} \delta_{c,d} \delta _{f,g} -4 \delta _{a,d} \delta _{b,c} \delta _{f,g}  -4  \delta _{a,c} \delta _{b,d} \delta _{f,g} ) \,.
\end{align}

Regarding the odd-numbered case, when the WZW interaction is present, there may exist additional mass-depletion processes, such as  $\pi\pi\pi \to \pi \pi$ and  $ \pi X  \to \pi \pi$. Concretely, the WZW interaction is given by
\begin{align}
\mathcal{L}_{\rm int}^{\rm odd}  =  &\frac{2 N_c}{15 \pi^2 f_\pi^5} \epsilon_{\mu \nu \rho \sigma} \operatorname{Tr}\left[\pi \partial^\mu \pi \partial^\nu \pi \partial^\rho \pi \partial^\sigma \pi\right]+ \dots \\
  =  &\frac{ N_{c}}{10 \sqrt{2} \pi^{2} f_{\pi}^{5}} \epsilon_{\mu \nu \rho \sigma} \left(\pi_1 \partial^{\mu}\pi_2 \partial^{\nu} \pi_3 \partial^{\rho} \pi_4 \partial^{\sigma} \pi_5 + \pi_2 \partial^{\mu}\pi_3 \partial^{\nu} \pi_4 \partial^{\rho} \pi_5 \partial^{\sigma} \pi_1  + \pi_3 \partial^{\mu}\pi_4 \partial^{\nu} \pi_5 \partial^{\rho} \pi_1 \partial^{\sigma} \pi_2  \right. \notag \\ 
 &~\left. \qquad\qquad\qquad\qquad  + \pi_4 \partial^{\mu}\pi_5 \partial^{\nu} \pi_1 \partial^{\rho} \pi_2 \partial^{\sigma} \pi_3 +\pi_5 \partial^{\mu}\pi_1 \partial^{\nu} \pi_2 \partial^{\rho} \pi_3 \partial^{\sigma} \pi_4 \right)  +\dots  \,,
\end{align}
where the ellipses stand for higher order odd-numbered interactions. The Lagrangian above leads to non-vanishing mass-depletion processes if all involved $\pi$'s have different flavors.  In this case, its Feynman rule is given by 
\begin{align}
\Gamma_{\left\{p_1, a \right\}, \left\{p_2, b\right\}, \left\{p_3, c\right\} , \left\{p_4, d\right\}, \left\{p_5, e\right\} } = \frac{ N_{c}}{2 \sqrt{2} \pi^{2} f_{\pi}^{5}}   \epsilon_{\mu \nu \rho \sigma} p_2^\mu p_3^\nu p_4^\rho p_5^\sigma  \quad (a\neq b\neq c\neq d\neq e)\,. 
\end{align}
Again, all momenta are defined with their direction pointing towards the interaction vertex.

{Our considered processes include free $\pi$ interactions and those involving one or two bound states~$X$.  For the latter, we adopt the notation of \cite{Peskin:1995ev} and define the normalized~$X$ state at rest in terms of   
\begin{equation}
    |X \rangle =  {\sqrt{2M_X} \over 2 m_\pi} \int {d^3 \vec q \over (2\pi)^3} \tilde \psi (\vec q)   |\vec q, - \vec q \rangle   \,, 
\end{equation}
together with $\int {d^3 \vec q \over (2\pi)^3} |\tilde \psi (\vec q) |^2 =1$;  $\pm \vec q $ gives the momenta of two internal~$\pi$ states. In comparison to a free process, the corresponding interaction amplitude hence contains an additional factor of ${\sqrt{2M_X} \over  2 m_\pi}\,  \int {d^3 \vec q  \over (2\pi)^3}\,\tilde \psi (\vec q)  $ (or its conjugate) for each incoming (outgoing) bound state. This factor is in agreement with what is termed the ``instantaneous approximation'' in~\cite{Petraki:2015hla} obtained in the non-relativistic limit. 
In the notation of the latter work, for an incoming (outgoing) $\pi$ particle with $3$-momentum $\vec k$, its wave function is given by
$ \tilde \phi_{\rm free}(\vec q) = (2\pi)^3 \delta^3{(\vec k -\vec q)}   $
(or its  conjugate),  as we focus on  finite-range interactions in this work.  
}

\subsection{Bound state formation \boldmath$3\pi \to \pi X$}

\begin{figure}[t]
\begin{center}
\includegraphics[width=0.7\textwidth]{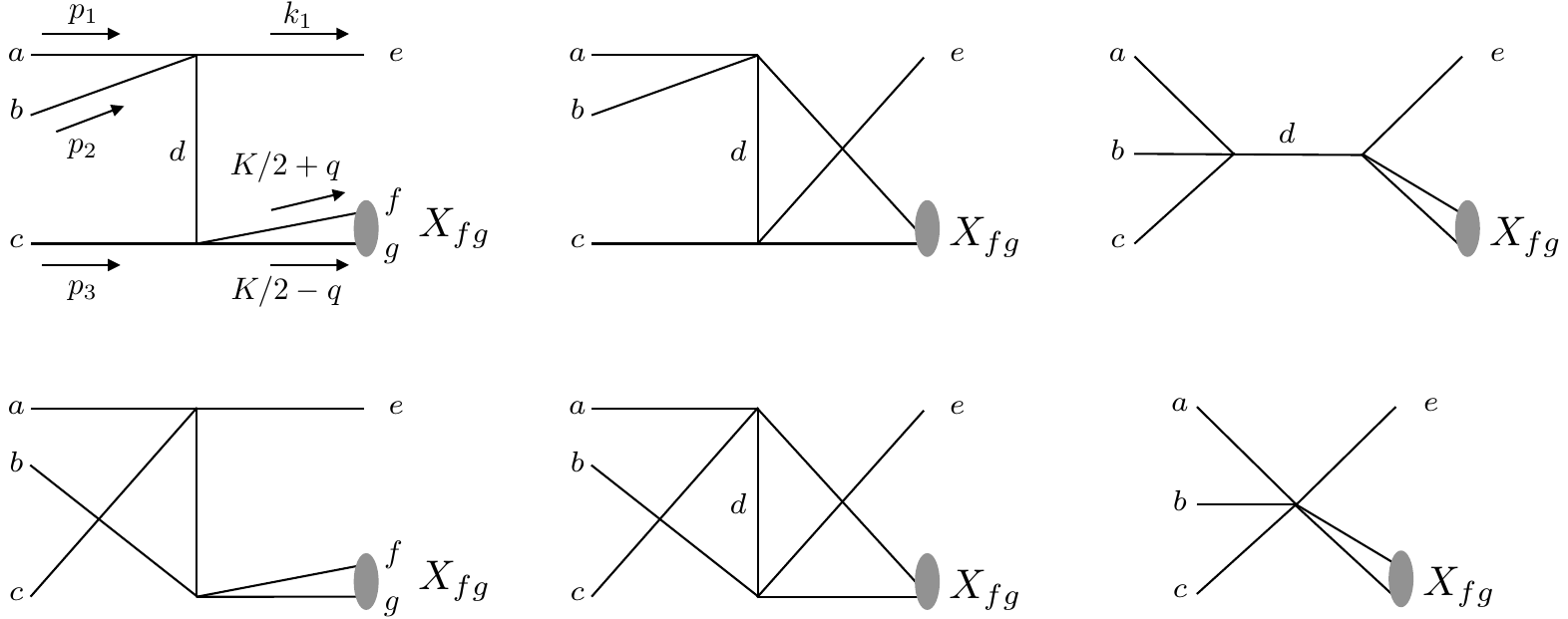}
\end{center}
\caption{Feynman diagrams contributing to $3\pi \to \pi X$. The $t$- and $u$-channel type diagrams (left and middle)  are enhanced  in the non-relativistic regime, relative to  $s$-channel (top left) and final 6-point contact interaction (bottom left) diagrams.}
\label{fig:Boltzmannw4P}
\end{figure}

With all Feynman rules at hand, we now compute the collision term for 
 bound state formation $3\pi \to \pi X$. The process is dominated by the first and second $t$- and $u$-channel diagrams of both rows of Fig.~\ref{fig:Boltzmannw4P}. For the first two, the amplitudes read,
\begin{align}
\mathcal M_t= & \frac{1}{k_t^2-m_\pi^2} \sum_{d} \int \frac{d^3 \vec q}{(2 \pi)^3}  \frac{\sqrt{2 M_X}}{2 m_\pi}  \widetilde{\psi}^*(\vec q)\,\Gamma_{\left\{a, p_1\right\},\left\{b, p_2\right\},\left\{e, - k_1\right\},\{d, - k_t\}}  \Gamma_{\{ d, k_t\},\left\{c, p_3\right\},\left\{f, -\frac{K}{2}- q\right\},\left\{g, - \frac{K}{2}+ q\right\}}  \,,\\
\mathcal M_u=  & \frac{1}{k_u^2-m_\pi^2} \sum_{d} \int \frac{d^3 \vec q}{(2 \pi)^3} \frac{\sqrt{2 M_X}}{2 m_\pi}\widetilde{\psi}^*(\vec q)\, \Gamma_{\left\{a, p_1\right\},\left\{b, p_2\right\},\left\{f, - \frac{K}{2}- q \right\},\{d, - k_u\}}   \Gamma_{\{ d, k_u\},\left\{c, p_3\right\},\left\{ e, - k_1 \right\},\left\{g, - \frac{K}{2}+ q\right\}}\,,
\end{align}
where $k_t = p_1 + p_2 - k_1$ and $k_u = p_1 + p_2 - \frac{K}{2} -q$ are the four-momenta of the $t$ and $u$-channel propagators, respectively.  In the non-relativistic limit and using 3-body kinematics in the center of momentum (CM) frame, it can be shown that $k_t^2-m_\pi^2  \simeq - 8 E_B m_\pi/3$ and $ k_u^2- m_\pi^2 \simeq  -E_B m_\pi/3$, respectively, leading to  near on-shell enhancements of both diagrams. In all simplifications we assume that  $E_B > T$ so that $m_\pi v^2/2E_B$ remains a small parameter where $v$ is a typical DM velocity, $v\sim \sqrt{T/m_\pi}$.  In addition, the wave-function of each bound state, assumed to be flavor-blind here,  is normalized as 
\begin{equation}
	\int \frac{d^3 q}{(2\pi)^3} \: \widetilde\psi^* (\vec q)   \widetilde\psi (\vec q) 
= \int d^3 r\   \psi^* (\vec r)   \psi (\vec r) = 1 \, ,\text{~~together with~~} \widetilde \psi ({\vec q}) \equiv \int d^3 r \ \psi ({\vec r})  e^{-i {\vec q \cdot \vec r} }\,.
\end{equation}
Thus, for constant terms of $\Gamma$, one can simplify the amplitude with the equality
\begin{equation}
 \int \frac{d^3 \vec q}{(2 \pi)^3} \frac{\sqrt{2 M_X}}{2 m_\pi}\widetilde{\psi}^*(\vec q) =  \int \frac{d^3 \vec q}{(2 \pi)^3} \frac{\sqrt{2 M_X}}{2 m_\pi}  \int d^3 r \psi^* ({\vec r})  e^{i {\vec q \cdot \vec r} } =  \frac{\sqrt{2 M_X}}{2 m_\pi}  \int d^3 r \psi^* ({\vec r})   \delta^3({\vec r})  =  \frac{\sqrt{2 M_X}  }{2 m_\pi}  \psi^* (0)\,.
\end{equation}
The $3\pi \rightarrow \pi X$ cross section of negative mass dimension five may then be written in the CM frame as
\begin{align}
 \sigma_{3 \pi \rightarrow \pi X} v^{2}  \, &\equiv  \frac{1}{2 E_{p_1} \, 2E_{p_2}\, 2 E_{p_3}}  \int \frac{d^3 k_1}{(2 \pi)^3 2 E_{k_1}} \frac{d^3 K}{(2 \pi)^3 2 E_{K}} (2 \pi)^4 \delta^4\left(p_1+p_2+p_3-k_1-K\right)|\mathcal{M}_{3\pi \rightarrow \pi X}|^{2} \label{eq:define3to2}\\
 &  \simeq   \frac{1}{128\pi^2 E_{p_1} E_{p_2} E_{p_3}} \frac{|\vec k_1|}{(E_{p_1} +E_{p_2}+E_{p_3})} \int d\Omega\,  |\mathcal{M}_{3\pi \rightarrow \pi X}|^2\,,
\end{align}
where $\mathcal{M}_{3\pi \rightarrow \pi X}$ is obtained from the sum of the four $t$- and $u$-channel-type diagrams of Fig.~\ref{fig:Boltzmannw4P} (the specific flavor combinations will be discussed momentarily);
$d\Omega$ is the differential final state solid angle.
The overall collision term or ``annihilation cross section''  is defined by the physical number-changing rate of free $\pi$ as it enters our Boltzmann equation in the previous section, i.e., in the absence of Hubble expansion
\begin{equation}
    {dn_\pi} = \sum_k {dn_{\pi_k}}=  2 n_{\pi}^3 \langle \sigma_{3\pi \rightarrow \pi X} v^{2}\rangle {dt} \, .
\end{equation}
By adding up all distinct processes that contribute to $dn_\pi/dt$, we obtain
\begin{align}
n_{\pi}^3 \langle \sigma_{3\pi \rightarrow \pi X} v^{2}\rangle  =  \sum_{a> b >c} n_{\pi_a}n_{\pi_b}n_{\pi_c} \langle \sigma_{\pi_a \pi_b\pi_c \rightarrow \pi X} v^{2}\rangle +  \sum_{a\neq b} {n_{\pi_a}^2\over 2!} n_{\pi_b} \langle \sigma_{\pi_a \pi_a \pi_b \rightarrow \pi X} v^{2}\rangle +  \sum_{a} {n_{\pi_a}^3 \over 3!} \langle \sigma_{\pi_a \pi_a\pi_a \rightarrow \pi X} v^{2}\rangle\,,
\end{align}
where the prefactors $2!$ and $3!$ avoid double-counting of identical initial states. As in the thermal bath $n_{\pi_a} = n_{\pi_b}=n_{\pi_c} =  n_\pi/N_\pi$ holds, our thermally-averaged cross section can be written as 
\begin{align}\label{eq:3pitopiX}
\langle \sigma_{3\pi \rightarrow \pi X} v^{2}\rangle ={1\over N_\pi^3}  {1\over 2!}\sum_{e,f,g}\left[  \sum_{a\neq  b \neq c}  {1\over 3!}\langle \sigma_{\pi_a \pi_b\pi_c \rightarrow \pi_e X_{fg}} v^{2}\rangle +  \sum_{a\neq b} {1\over 2!}   \langle \sigma_{\pi_a \pi_a \pi_b \rightarrow \pi_e X_{fg}} v^{2}\rangle +  \sum_{a} {1 \over 3!} \langle \sigma_{\pi_a \pi_a\pi_a \rightarrow \pi_e X_{fg}} v^{2}\rangle\,\right]\,,
\end{align}
where $N_\pi=5$, 
and $\frac{1}{2!}\sum_{e,f,g}$
sums up the final 
states, taking into account  that $X_{fg} =X_{gf}$. For the specific flavor combinations, the following matrix elements enter
\begin{align}
    \mathcal{M}_{\pi_a \pi_b\pi_c \rightarrow \pi_e X_{fg}} & =    (\mathcal M_{t} +  \mathcal M_{u} )  +  ({\left\{a, p_1\right\} \leftrightarrow \left\{c, p_3\right\} } ) +  ({\left\{b, p_2\right\} \leftrightarrow \left\{c, p_3\right\}})\,,\\
        \mathcal{M}_{\pi_a \pi_a\pi_c \rightarrow \pi_e X_{fg}} & =     (\mathcal M_{t} +  \mathcal M_{u} )|_{b\to a} +  ({\left\{a, p_1\right\} \leftrightarrow \left\{c, p_3\right\} } )|_{b\to a}\,,\\
            \mathcal{M}_{\pi_a \pi_a\pi_a \rightarrow \pi_e X_{fg}} & =     (\mathcal M_{t} +  \mathcal M_{u} )|_{b\to a,\,c\to a} \,,
\end{align}
where $\mathcal M_{t,u}$ have been defined above, and $({\left\{...\right\} \leftrightarrow \left\{...\right\}})$ stands for switching the corresponding two particles in the first term on the right hand side (RHS) of each corresponding equation. Putting everything together, we obtain the effective cross section for the bound state formation rate as
\begin{equation}
   \langle \sigma_{3\pi \rightarrow \pi X} v^{2}\rangle \simeq \frac{\numprint{57041}   |\psi(0)|^2 }{\numprint{327680} \sqrt{3} \pi   f_\pi^8 } \left({ m_\pi \over E_B} \right)^{3/2}  +  O\left(\sqrt{ m_\pi \over E_B} \right) =   \frac{\numprint{57041}   R(0)^2 }{\numprint{1310720} \sqrt{3} \pi^2   f_\pi^8  }\left({ m_\pi \over E_B} \right)^{3/2}  +  O\left(\sqrt{ m_\pi \over E_B} \right)\,.
\end{equation}
This equation is applicable in the non-relativistic limit, where the ground-state wavefunction of the two-particle bound state is $\psi(r) = R(r)/\sqrt{4\pi}$. The numerical prefactor in the last equality is approximately~$3\times 10^{-3}$. Note that we have neglected the leftmost diagrams of Fig.~\ref{fig:Boltzmannw4P}, which are not enhanced by inverse powers of binding energy.  
 
\begin{figure}[t]
\begin{center}
\includegraphics[width=\textwidth]{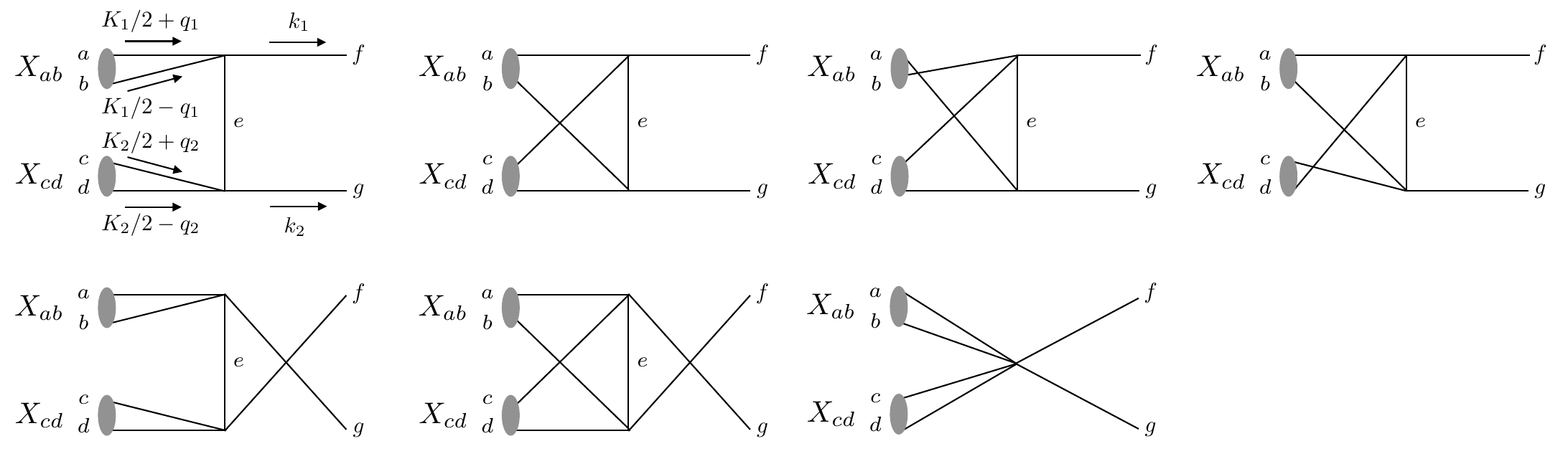}
\end{center}
\caption{Feynman diagrams contributing to $XX \to \pi\pi$. Some of the diagrams are identical when $a = b$ or~$c=d$. }
\label{fig:Boltzmannw6P}
\end{figure}

\subsection{{\boldmath$XX \to \pi\pi$ annihilation and free $4\pi \to \pi \pi$ process}}

Now, we turn to the mass-depleting process $XX\to \pi\pi$ mediated by  even-numbered interactions and additionally provide the formulae for the free $4\pi \to 2\pi$ at the non-relativistic limit.
For $XX\to \pi\pi$, we take into account all diagrams shown in Fig.~\ref{fig:Boltzmannw6P}, among which the amplitude of the first one is given by
\begin{equation}
 \mathcal M_{abf,cdg}=  \frac{1}{k^2-m_\pi^2} \sum_{e} \int \frac{d^3 \vec q_1 d^3 \vec q_2 }{(2 \pi)^6} \frac{2 M_X \widetilde{\psi}(\vec q_1) \widetilde{\psi}(\vec q_2) }{(2 m_\pi)^2}  \,\Gamma_{\left\{a, {K_1 \over 2} +q_1\right\},\left\{b, {K_1 \over 2}  - q_1 \right\},\left\{f,  -k_1\right\},\{e, -k\}}  \Gamma_{\left\{c, {K_2 \over 2} +q_2\right\},\left\{d, {K_2 \over 2} - q_2\right\},\left\{e, k \right\},\left\{g,  -k_2\right\}}  \,,
\end{equation}
where the subscripts $abf$ and $cdg$ denote the incoming and outgoing states that are connected by the same vertex. The  4-momentum of the intermediate state is given by $k = (K_1/2 +q_1) +  (K_1/2 -q_1) - k_1 = K_1 - k_1$, pointing from the  $f$- to the $g$-associated vertex. The amplitudes of the following five diagrams can be  obtained by switching the momenta and indices in both $\Gamma$'s and $k$, labelled as $ \mathcal M_{acf,bdg}$, $ \mathcal M_{bcf,adg}$, $ \mathcal M_{adf,bcg}$, $ \mathcal M_{cdf,abg}$, $ \mathcal M_{bdf,acg}$, respectively.  The amplitude of the last one,  $\mathcal M_{abcdfg}$, is the  sum of $\Gamma_{\rm 6-point}|_{\rm contact}$ and $\Gamma_{\rm 6-point}|_{\rm kinetic}$ that have been given above. 

Again we define the thermally-averaged cross section such that it yields the physical mass-depletion rate, i.e.,  $ {dn_X}/{dt} =  2 n_X^2 \langle \sigma_{2X \rightarrow 2 \pi} v\rangle$ in the absence of Hubble expansion. Here, $n_\pi$ and $n_X$ sum up all flavors  of free and bound states. Denoting by $N_X = N_\pi(N_\pi + 1)/2 = 15$ the number of flavor-blind bound states we have
\begin{align}\label{eq:massdeplete6}
\langle \sigma_{2X \rightarrow 2\pi} v\rangle = & {1\over N_X^2}  \sum_{f > g}  \left[  \sum_{a =   c > b > d}    \langle \sigma_{X_{ab} X_{cd} \rightarrow \pi_f \pi_g}  v\rangle  +  \sum_{b> a =  c  > d}   \langle \sigma_{X_{ab} X_{cd} \rightarrow \pi_f \pi_g}  v\rangle   +  \sum_{b > d > a =  c}  \langle \sigma_{X_{ab} X_{cd} \rightarrow \pi_f \pi_g}  v\rangle   \right]\notag \\
& +  {1\over N_X^2}  \sum_{f > g}  \left[  \sum_{a =   b > c > d}    \langle \sigma_{X_{ab} X_{cd} \rightarrow \pi_f \pi_g}  v\rangle  +  \sum_{c> a =   b  > d}   \langle \sigma_{X_{ab} X_{cd} \rightarrow \pi_f \pi_g}  v\rangle   +  \sum_{c > d > a =   b }  \langle \sigma_{X_{ab} X_{cd} \rightarrow \pi_f \pi_g}  v\rangle   \right]\notag \\
 & + {1\over N_X^2}  \sum_{f > g }   \left[   \sum_{a = b = c >   d  }  \langle \sigma_{X_{ab} X_{cd} \rightarrow \pi_f \pi_g}  v\rangle  +   \sum_{d > a = b = c  }  \langle \sigma_{X_{ab} X_{cd} \rightarrow \pi_f \pi_g}  v\rangle    \right] \notag \\
 & + {1\over N_X^2}  \sum_{f}  \left[ \sum_{a = b >  c =  d  }   \langle \sigma_{X_{ab} X_{cd} \rightarrow \pi_f \pi_f}  v\rangle  +   {1\over 2! }  \sum_{a = b = c =  d  }   \langle \sigma_{X_{ab} X_{cd} \rightarrow \pi_f \pi_f}  v\rangle  \right]  \,, 
\end{align}
where there are no $f =  g$ ($f > g$) terms in the first three lines (last line), and $a,\,b,\,c,\,d$ cannot be all different from each other,  because  all even-numbered interaction are   flavor-conserving  in this concrete model. In fact, the final states are uniquely fixed by the flavors of initial states once  $f\ge g$ is chosen.   In the last line, a factor of $1/2!$ avoids double-counting  of the identical initial states. Meanwhile, the scattering angle ranges only from $0$ to $\pi/2$ for identical final states.
Finally, note that $ \sum_{a =   c > b > d}    +  \sum_{b> a =  c  > d}  +  \sum_{b > d > a =  c} $ in the first line can be unified to $\sum_{a =   c \neq b \neq d} /2$. However, it then becomes less obvious that there is no double-counting issue with the latter expressions. A similar simplification applies to the second and third lines. For instance, the third line corresponds to $\sum_{a =  b = c \neq d} $.

As shown in Fig.~\ref{fig:Boltzmannw6P}, some diagrams are the same when $a = b$, as the effect of switching two identical particles in a bound state is absorbed by the corresponding wavefunction.   Moreover,  when $a=b$ and $c=d$, only the first, second, fifth, and last diagrams are distinct. Taking this into account, we  express the general  amplitude for Eq.~\eqref{eq:massdeplete6} as  
\begin{equation}
  \mathcal M_{X_{ab}X_{cd}\to \pi_e \pi_f}  =  \mathcal M_{abf,cdg} + \mathcal M_{acf,bdg} + (1-\delta_{a,b})\mathcal M_{bcf,adg} +  (1-\delta_{c,d})  \mathcal M_{adf,bcg}+\mathcal M_{cdf,abg} + (1-\delta_{a,b}) \mathcal M_{bdf,acg} + \mathcal M_{abcdfg}\,.  \notag 
\end{equation}
With this amplitude,  the effective cross section for the mass-depletion rate can be calcualted to yield 
\begin{align}
   \langle \sigma_{2X \rightarrow 2\pi} v \rangle \simeq & \frac{|\psi(0)|^4}{  f_\pi^8 }  \sqrt{1-{m_\pi^2 \over  m_X^2 }} %
   \left(\frac{\numprint{118790261}\,  m_X^8  -\numprint{104839888} \,m_\pi^2 m_X^6  +\numprint{36685056}\, m_\pi^4 m_X^4     -\numprint{1573120}\, m_\pi^6 m_X^2+ \numprint{4444800}\,  m_\pi^8}{\numprint{23887872000}\, \pi \,  m_\pi^4 m_X^4} \right)\notag\\ 
  \simeq  & \frac{\numprint{2529757}}{\numprint{26542080} \sqrt{3} \pi  }  \frac{|\psi(0)|^4}{f_\pi^8 } = \frac{\numprint{2529757}}{\numprint{424673280}\sqrt{3} \pi^3}  \frac{R(0)^4}{f_\pi^8 }    \,
\end{align}
in the non-relativistic limit, where we have taken $m_X \simeq 2m_\pi $ and $\psi(0)=R(0)/\sqrt{4\pi}$ for the last two equalities. The numerical prefactor in the last equality is~$10^{-4}$.

{
The calculation of the free $4\pi\to 2\pi$ cross-section proceeds in a similar fashion. Again, here we define the effective cross section through the total $\pi$-number changing rate, i.e., $dn_\pi = 2 n_\pi^4 \langle  \sigma_{4\pi \rightarrow 2\pi} v^3  \rangle dt$ (neglecting the Hubble rate). We then obtain 
\begin{align}\label{eq:count4to2}
\langle \sigma_{4\pi \rightarrow 2\pi} v^3\rangle =  {1\over N_\pi^4}  \sum_{f > g} &   \left[{1\over 2!} \sum_{a =   b >  c >  d}    \langle \sigma_{\pi_a \pi_b \pi_c \pi_d  \rightarrow \pi_f \pi_g}  v^3\rangle  + {1\over 2!} \sum_{a  > b =  c > d}  \langle \sigma_{\pi_a \pi_b \pi_c \pi_d  \rightarrow \pi_f \pi_g}  v^3 \rangle     + {1\over 2!} \sum_{a > b > c =  d}   \langle \sigma_{\pi_a \pi_b \pi_c \pi_d  \rightarrow \pi_f \pi_g}  v^3\rangle     \right] \notag \\
 & + {1\over N_\pi^4}  \sum_{f > g}  \left[ {1\over 3!}\sum_{a =   b =  c >  d}    \langle \sigma_{\pi_a \pi_b \pi_c \pi_d  \rightarrow \pi_f \pi_g}  v^3\rangle  +   {1\over 3!} \sum_{a > b = c = d }  \langle \sigma_{\pi_a \pi_b \pi_c \pi_d  \rightarrow \pi_f \pi_g}  v^3\rangle    \right] \notag \\
 & + {1\over N_\pi^4}   \sum_{f=g}  \left[ {1\over 2!} {1\over 2!}\sum_{a = b >  c =  d  }   \langle \sigma_{\pi_a \pi_b \pi_c \pi_d    \rightarrow \pi_f \pi_g}  v^3\rangle  +   {1\over 4! }  \sum_{a = b = c =  d  }   \langle \sigma_{\pi_a \pi_b \pi_c \pi_d   \rightarrow \pi_f \pi_g}  v^3 \rangle  \right]  \,, 
\end{align}
after counting all non-equivalent, non-vanishing combinations of  flavors on the right hand side of the equation, where the prefactors have been added to avoid double-counting identical initial states. Note that the last line produces identical final states. Here we take the non-relativistic limit, so all initial states have the same $4$-momentum ($m_\pi$, 0,0,0),  resulting in
\begin{align}\label{eq:sigmav4to2}
\langle \sigma_{4\pi \rightarrow 2\pi} v^3 \rangle  = \, & {\sqrt{3} \over 256 \pi m_\pi^4 } |{ \mathcal M}_{4\pi \rightarrow 2\pi} |^2   \simeq {\numprint{10432975} \over \numprint{7077888} \sqrt{3} \pi N_\pi^4}\,\frac{1}{  f_\pi^8}\,.
\end{align}
where we have adopted the definition of the four-body cross section from \cite{Bhattacharya:2019mmy}. The prefactor is approximately $4\times 10^{-4}$ after adopting the actual flavor number~$N_\pi =5$. 
}

\subsection{Standard \boldmath$3\pi \to 2\pi$ and catalyzed \boldmath$\pi X \to \pi\pi$ process}

The final set of cross sections used in this work are the ones induced by the WZW term. Both, the free $3\pi \to 2\pi$ and its bound-state version $\pi X\to \pi\pi$,  are velocity-dependent in the non-relativistic freeze-out regime. We therefore have to include momentum-dependent terms in the amplitudes. The case  involving only free states, $3\pi \to  2\pi$, is simple, as its amplitude is directly given by the vertex factor, 
\begin{equation}
\mathcal M_{\pi_a  \pi_b   \pi_c \to \pi_d \pi_e}= \Gamma_{\left\{p_1, a \right\}, \left\{p_2, b\right\}, \left\{p_3, c\right\} , \left\{-k_1, d\right\}, \left\{-k_2, e\right\} }  \,,
\end{equation}
where $p_i$ and $k_i$ are the momenta of incoming and outgoing states, respectively. Taking the definition of a $3\to2$ cross section above, Eq.~\eqref{eq:define3to2},  one obtains in the CM frame 
\begin{align}\label{eq:3to2M}
    \sigma_{\pi_a   \pi_b    \pi_c \to \pi_d \pi_e } v^{2}  &=   \frac{1}{128 \pi^2  E_{p_1} E_{p_2} E_{p_3}} \frac{|\vec{k}_{1}|}{\sqrt{s}} \int  d\Omega \,  |\mathcal M_{\pi_a  \pi_b    \pi_c \to \pi_d \pi_e}|^{2}  \notag \\&
    =\frac{ N_c^2    }{3072 \pi^5 f_\pi^{10} } \frac{ s (s-4m_\pi^2)^{3/2}   \left[s_{12} s_{13} s_{23} - m_\pi^2 (s - m_\pi^2)^2\right] }{(s+m_\pi^2 -s_{12})(s+m_\pi^2 -s_{13})(s+m_\pi^2 -s_{23})}\,,
\end{align}
where $s=(p_1 + p_2 + p_3)^2$, $s_{ij}=(p_i + p_j)^2$. Note that $a,b,c,d,e$ are all different.  The factor $ \left[s_{12} s_{13} s_{23} - m_\pi^2 (s - m_\pi^2)^2\right] $ vanishes in the limit $s_{ij} \to 4m_\pi^2$ and  $s \to 9m_\pi^2$. Therefore, one requires an expansion of the Mandelstam variables~$s$ and $s_{ij}$ to higher order in velocities. 

Without including relativistic effects, we calculate the thermally-averaged  cross section as follows: 
\begin{equation}\label{eq:thermalInt}
     \left\langle \sigma_{\pi_a   \pi_b    \pi_c \to \pi_d \pi_e } v^{2}  \right\rangle= \frac{\int d^{3} \vec  p_{1} d^{3} \vec  p_{2} d^{3} \vec  p_{3} \, f^{\rm eq}(\vec p_1) f^{\rm eq}(\vec p_2) f^{\rm eq}(\vec p_3) \times ( \sigma_{\pi_a   \pi_b    \pi_c \to \pi_d \pi_e } v^{2}  )    }{\int d^{3} \vec p_{1} d^{3} \vec  p_{2} d^{3} \vec  p_{3}  \, f^{\rm eq}(\vec p_1) f^{\rm eq}(\vec p_2) f^{\rm eq}(\vec p_3)  }\,,
\end{equation}
where we adopt non-relativistic Maxwell-Boltzmann equilibrium statistics for the three initial states distribution functions~$f^{\rm eq}$.  
To simplify this integral, the key is that for each set of ($\vec p_1, \,\vec p_2,\, \vec p_3$) we have the freedom to choose a concrete frame where we fix the $y$-$z$ plane with $\vec p_1, \,\vec p_2$, and set $\vec{p}_1$ along the z-direction,  so that
\begin{align}
    &p_1 = \left(E_{p_1},\, 0,\,0,\, |\vec{p}_1| \right) \,,\\&
    p_2 = \left(E_{p_2}, \, 0,\, |\vec{p}_2| \sin \theta_{2}, \, |\vec{p}_2|\cos \theta_{2} \right) \,,\\&
    p_3 = \left(E_{p_3}, \, |\vec{p}_3|\sin \theta_{3} \cos \phi_{3}, \, |\vec{p}_3|\sin \theta_{3}  \sin\phi_{3}, \, |\vec{p}_3| \cos \theta_3\right)\,,
\end{align}
where $E_{p_i} = \sqrt{p_i^2 + m_\pi^2 }$, $\theta_{2}$ is the angle between  ($\vec p_1, \,\vec p_2$), independent of the frame after neglecting second-order relativistic corrections, and  the direction of $\vec p_3$ is randomly distributed.  That is, the distribution functions of all six variables, ($|\vec{p}_1| $, $|\vec{p}_2| $, $|\vec{p}_3| $, $\theta_{2}$, $\theta_{3}$,  $\phi_{3}$), are fixed independently by the three-dimensional Maxwell-Boltzmann distribution functions.  

Those six variables can then describe the squared amplitude, in which the Lorentz-invariant Mandelstam variables become
\begin{align}
    & s_{12}= 2 m_\pi^2+ 2 E_{p_1} E_{p_2}- 2 \sqrt{E_{p_1}^2- m_\pi^2}\sqrt{E_{p_2}^2- m_\pi^2} \cos \theta_2 \,,\\&
    s_{13}= 2 m_\pi^2+ 2E_{p_1}E_{p_3}- 2 \sqrt{E_{p_1}^2- m_\pi^2}\sqrt{E_{p_3}^2- m_\pi^2}  \cos \theta_3 \,,\\&
    s_{23}= 2 m_\pi^2+ 2 E_{p_2}E_{p_3}- 2 \sqrt{E_{p_2}^2- m_\pi^2}\sqrt{E_{p_3}^2- m_\pi^2}  \left(\cos \theta_2\cos \theta_3+  \sin \theta_2 \sin \theta_3 \sin \phi_3\right)\,, \\&
    s = s_{12}+s_{13}+ s_{23}-3 m_\pi^2\,. 
\end{align}
As a result, the momentum-dependent part of the integral in  Eq.~\eqref{eq:thermalInt} now simplifies to 
\begin{equation}
\int d^{3} \vec  p_{1} d^{3} \vec  p_{2} d^{3} \vec  p_{3}   =   \int_0^\infty 4\pi |\vec{p}_1|^2\, d|\vec{p}_1|  \int_0^\infty 2\pi |\vec{p}_2|^2\,  d|\vec{p}_2|  \int^{1}_{-1}   d\cos\theta_{2} \int_0^\infty |\vec{p}_3|^2  d|\vec{p}_3|   \int^{1}_{-1} d\cos\theta_{3} \int^{2\pi }_{0} d\phi_{3} \,, 
\end{equation}
which  can be solved analytically. We obtain
\begin{equation} 
     \left\langle \sigma_{\pi_a   \pi_b    \pi_c \to \pi_d \pi_e } v^{2}  \right\rangle =  \frac{5 \sqrt{5} N_c^2   m_\pi^3 T^2   }{1024 \pi^5 f_\pi^{10} }  +  O\left({ T^4\over m_\pi^4} \right)\,,
\end{equation}
where $T$ is the temperature of the dark sector.  
Since this cross section is only non-vanishing for  $a\neq b\neq c$, the additional prefactor after averaging over all flavors of initial states is the same as the first term on the RHS of Eq.~\eqref{eq:3pitopiX}, resulting in 
\begin{equation} \label{eq:sigmav3to2}
\langle \sigma_{3\pi \rightarrow 2\pi} v^{2} \rangle = {1\over N_\pi^3 }{1\over 2! \, 3!} \sum_{a \neq b \neq c \neq d \neq e} \langle  \sigma_{\pi_a   \pi_b   \pi_c \to \pi_d \pi_e} v^{2} \rangle   = {1 \over N_\pi^3 }{5! \over 2!\, 3!} \frac{5 \sqrt{5}N_c^2     m_\pi^3 T^2   }{1024 \pi^5 f_\pi^{10} }  \,,
\end{equation}
where the new factor $5!$ is the number of non-equal ($a,b,c,d,e$) combinations. The numerical factor in the last expression is approximately $10^{-5}$ with $N_c =2$. This result has an additional factor of $1/3$ with respect to the thermally averaged cross section in~\cite{Hochberg:2014kqa} after taking into account the different normalizations $f_\pi = f_\pi^\text{Ref.\cite{Hochberg:2014kqa}}/2$ and $T^n = T^n_\text{Ref.\cite{Hochberg:2014kqa}} /2 $,  but is in agreement with~\cite{Kamada:2022zwb}.

Similarly, the mass-depletion process involving the bound state can be calculated from the squared amplitude  as
\begin{align}
  &   |\mathcal M_{\pi_a X_{bc} \to \pi_d \pi_e}|^2 = \bigg| \int \frac{d^3 \vec q }{(2 \pi)^3}\frac{\sqrt{2 M_X} \widetilde{\psi}^*(\vec q) }{2 m_\pi } \Gamma_{\left\{p, a \right\}, \left\{P/2 +q , b\right\}, \left\{P/2  - q, c\right\} , \left\{-k_1, d\right\}, \left\{-k_2, e\right\} } \bigg|^2 \notag \\ 
 & \quad\quad\quad  \simeq   \frac{N_c^2  (s-4m_\pi^2) \left[m_X^4 + (s-4m_\pi^2)^2 - 2m_X^2(s+ 4m_\pi^2)^2 \right] \cos\theta_k^2  }{32\pi^4 f_\pi^{10} }  \left( {2m_X \over 4m_\pi^2 } \,  \bigg| \int \frac{d^3 \vec q  }{(2 \pi)^3}  \widetilde{\psi}^*(\vec q)   {q_z }  \bigg|^2 \right)   \,,
\end{align}
in the CM frame,  where  the $z$ direction is defined to be perpendicular to the plane spanned by the two vectors  $\vec p$ and  $\vec k_1$; the angle between the two latter 3-momenta is the scattering angle~$\theta_k$. We have set $a\neq b \neq c \neq d \neq e$, as the cross section vanishes for all other combinations.  The leading contribution contains $q_z$, requiring to solve
\begin{equation}
 \int \frac{d^3 \vec q }{(2 \pi)^3}  \widetilde{\psi}^*(\vec q)  q_z  =  \int {d^3 r}\psi^* (\vec r) \int \frac{d^3 q}{(2\pi)^3} \, e^{i  \vec q \cdot \vec r }  { q_z }  =  \int {d^3 r}\psi^*  (\vec r)  \left(-i {\partial \over \partial z}  \int \frac{d^3 q}{(2\pi)^3} \, e^{i  \vec q \cdot  \vec r }\right)  = -i  {\partial \over \partial z}   \psi^*  (\vec r) |_{\vec r \to 0} \,.
\end{equation}
Here, the first non-vanishing contribution comes from the $P$-wave bound state with wavefunctions in the form of $\psi_{1m}(\vec r) = Y_{1m}R(r)$, where $m =0,\pm 1$. After averaging over the value of $m$ and using $R(r)/r|_{r\to 0} =  R'(0) $, we obtain 
\begin{equation}\label{eq:k1wavefunction}
{1\over 3 }\sum_{m = 0,\pm 1 } \left| \int \frac{d^3 \vec q  }{(2 \pi)^3}  \widetilde{\psi}_{1m}^*(\vec q)   {q_z }  \right|^2  = {1\over 4\pi }R'(0)^2 \,.
\end{equation}
In addition, we have also calculated this quantity without fixing a predefined $z$-direction, and reach 
\begin{align}
    \frac{1}{2\ell + 1} \sum_m ( \nabla_i\psi^*_{\ell m})( \nabla_j\psi_{\ell m} ) \stackrel{\ell=1}{=} 
    \frac{1}{4\pi}
    \begin{pmatrix}
      R'(0)^2& 0 & 0 \\  
      0  & R'(0)^2  & 0 \\  
      0  & 0 & R'(0)^2  \\ 
    \end{pmatrix}_{\text{Cartesian tensor}} \,,
\end{align}
where $i,j = x,y,z$, and it agrees with Eq.~\eqref{eq:k1wavefunction}. With this at hand, we may now express the $2\to2$ cross section as
\begin{align}
\sigma_{\pi_a X_{bc} \to \pi_d \pi_e} v = &   \frac{|\vec k_1| }{64 \pi^2 E_p E_P (E_p + E_P)} \int d\Omega  |\mathcal M_{\pi_a X_{bc} \to \pi_d \pi_e}|^2 \notag\\
\simeq  & \frac{N_c^2 R'(0)^2 m_X s^2}{6144 \pi^6 f_\pi^{10} m_\pi^2 }\left[\frac{m_X^4 + (s-m_\pi^2)^2 - 2 m_X^2 (s+ m_\pi^2) }{ s^2 - (m_X^2 -m_\pi^2 )^2 }  \right] \left(1-\frac{4m_\pi^2}{s}\right)^\frac{3}{2}\,.
\end{align}
In the non-relativistic limit, we take the expansion $s= (m_\pi + m_X)^2+ m_\pi m_X v^2 +\mathcal O(v^4)$ in terms of the initial relative velocity $v$, leading to
\begin{align}
\sigma_{\pi_a X_{bc} \to \pi_d \pi_e} v = & \frac{N_c^2 R'(0)^2   (m_X^2+2m_X m_\pi  - 3m_\pi^2 )^\frac{3}{2} m_X^2 }{6144 \pi^6 f_\pi^{10} m_\pi (m_X+m_\pi) }   v^2 \,.
\end{align}
Finally, the thermally-averaged cross section obtained from the requirement $ {dn_X}/{dt} =  n_\pi  n_X \langle \sigma_{\pi X \rightarrow \pi \pi} v\rangle$ (neglecting Hubble expansion) is given by
\begin{align}
\langle \sigma_{\pi X \rightarrow 2\pi} v \rangle =&  {1\over N_\pi N_X}{5! \over 2! \, 2!} \langle  \sigma_{\pi_a X_{bc}    \to \pi_d \pi_e} v \rangle \notag\\
 = &  {1\over N_\pi N_X}{5! \over 2! \, 2!}   \frac{N_c^2 R'(0)^2   (m_X^2+2m_X m_\pi  - 3m_\pi^2 )^\frac{3}{2} m_X^2 }{6144 \pi^6 f_\pi^{10} m_\pi (m_X+m_\pi) }   \langle v^2  \rangle  \notag\\
\simeq  & \frac{N_c^2 R'(0)^2  (m_X^2+2m_X m_\pi  - 3m_\pi^2 )^\frac{3}{2}m_X }{5120 \pi^6 f_\pi^{10} m_\pi^2 }  T \,\notag\\
 = & \frac{\sqrt{5}N_c^2 R'(0)^2  m_\pi^2 }{512 \pi^6 f_\pi^{10}  }  T \,,
\end{align}
where in the first equality we include the following symmetry factors: $5!$ for the non-equal flavor combinations of the process, $1/2!$ accounting for  $X_{bc}  = X_{cb}$ and another $1/2!$ for the final states. For the next-to-last equality we use  $\langle v^2  \rangle = 3T(m_X^{-1}+m_\pi^{-1})$ and in the last equality we make the approximation $m_X\simeq 2m_\pi$. Setting $N_c =2$ yields for the numerical prefactor~$2\times 10^{-5}$.

\section{Additional details on abundance evolution}

In the main text we adopt a benchmark model fixing $m_\pi =  5f_\pi = 0.2$\,GeV, as well as  $R(0)=0.3m_\pi^{3/2}$, $dR(0)/dr=0.03m_\pi^{5/2}$ to demonstrate our numerical solutions to the abundance evolution as a function of $x = m_\pi/T$.  Here we provide additional quantitative information and study the  dependence on various parameters.

\subsection{Standard cases without bound states}

{In the absence of bound states~$X$ and under the assumption that the dark sector is in kinetic equilibrium with the photon bath the evolution of $Y_\pi$ simplifies to 
\begin{equation}
    {dY_\pi \over dx } = \frac{1}{3 H(x)}\frac{ds}{dx}
\left[ 
s \langle\sigma_{3\pi\to 2\pi} v^2\rangle  \left(  Y_{\pi}^3 -Y_{\pi}^2 Y_{\pi}^{\rm eq} \right)
+
2 s^2 \langle\sigma_{4\pi\to 2\pi} v^3\rangle  \left(  Y_{\pi}^4 -Y_{\pi}^2 (Y_{\pi}^{\rm eq})^2 \right)
\right] \,.
\end{equation}
While we also solve this case numerically in the main text, the required cross section to yield the observed relic abundance can be estimated analytically by setting the annihilation rate equal to the Hubble rate at $x_{\rm f}\sim 15\text{-}25$. 
}

{
If there are only even-numbered interactions, the condition $2\langle\sigma_{4\pi\to 2\pi} v^3\rangle  s(x_f)^3  Y_{\pi}^3 = H(x_f)$ together with the observed DM yield $Y_{\rm obs} \simeq 4 \times 10^{-10}\,\text{GeV}/m_\pi$ suggest  
\begin{equation}
  \Omega^{4\to 2}_{\pi} =  0.2 \left( {Y_{\pi} \over Y_{\rm obs} } \right) \sim  150\left( {10^8\,{\rm GeV}^{-8}  \over \langle\sigma_{4\pi\to 2\pi} v^3\rangle}\right)^{\frac{1}{3}} \left( {x_f \over  20} \right)^{\frac{7}{3} }  \left( {\text{GeV}\over  m_\pi } \right)^{\frac{4}{3} } \,  .
\end{equation}
Substituting $m_\pi =  5f_\pi = 0.2$\,GeV and $x_{\rm f} \sim 17$ into~\eqref{eq:sigmav4to2} results in $ \Omega^{4\to 2}_{\pi} \sim 10^3$, being much larger than the abundance of our benchmark model with bound state formation.   
}

{
In turn, when there exist dominant odd-numbered interactions, the condition $\langle\sigma_{3\pi\to 2\pi} v^2 \rangle s(x_f)^2  Y_{\pi}^2 = H(x_f)$  leads to    
\begin{equation}
  \Omega^{3\to 2}_{\pi} = 0.2 \left( {Y_{\pi} \over Y_{\rm obs} } \right) \sim    \left( {10^3\,{\rm GeV}^{-5}  \over \langle\sigma_{3\pi\to 2\pi} v^3\rangle}\right)^{\frac{1}{3}} \left( {x_f \over  21} \right)^2   \left( {\text{GeV}\over  m_\pi } \right)^{\frac{1}{2} } \,  .
\end{equation}
Again, taking~\eqref{eq:sigmav3to2} with $m_\pi =  5f_\pi = 0.2$\,GeV and $x_{\rm f} \sim 18$  gives $ \Omega^{3\to 2}_{\pi} \sim 10$. Although this value is significantly smaller than for the free $4\pi\to 2\pi$ freeze-out value above, it still over-closes the Universe. In contrast, for the same couplings, the observed relic abundance can be achieved with the assistance of the bound state, as is shown in the main text and below.  }

\subsection{Even-numbered case: analytical and numerical solutions} 

\begin{figure}[t]
\includegraphics[width=0.33\textwidth]{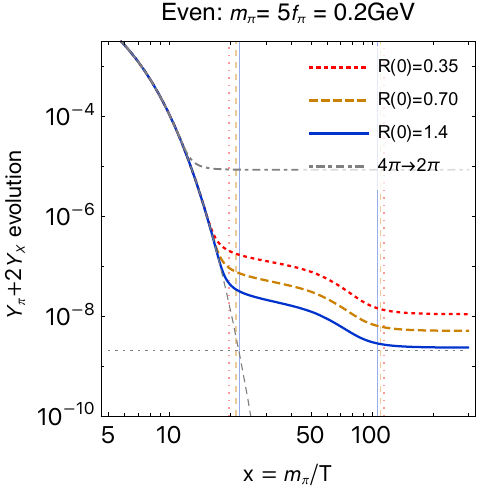}%
\hfill
\includegraphics[width=0.33\textwidth]{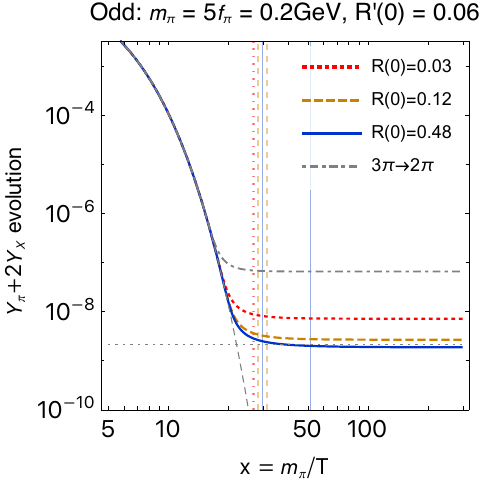}%
\hfill
\includegraphics[width=0.33\textwidth]{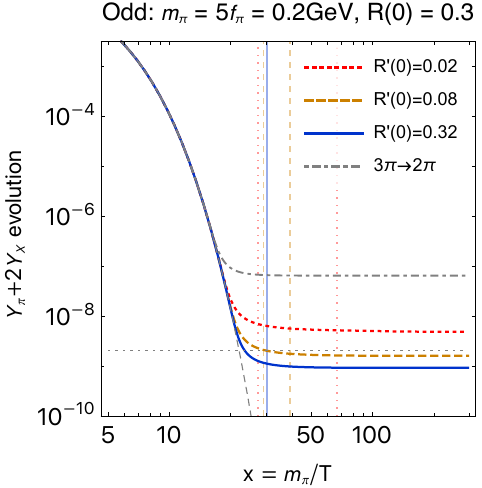}%
\caption{
Evolution of DM abundance, $Y_{\pi} + 2Y_{X}\simeq Y_{\pi} $ with $m_\pi = 5f_\pi =  0.2$\,GeV and  binding energy $\kappa_S =4\kappa_P = 0.1$. \emph{Left}: even-numbered cases varying $R(0)/ m_\pi^{3/2} = 0.35,\, 0.70,\, 1.4$ (from top to bottom). \emph{Middle}: odd-numbered cases varying $R(0)/ m_\pi^{3/2} = 0.03,\, 0.12,\, 0.48$ (from top to bottom) and fixing $R'(0)/ m_\pi^{3/2} = 0.06$.  \emph{Right}: odd-numbered cases fixing $R(0)/ m_\pi^{3/2} = 0.3$ and varying $R'(0)/ m_\pi^{3/2} = 0.02,\,0,08,\, 0.32$ (from top to bottom). The gray curves labeled ``$4\pi \to 2\pi$'' and ``$3\pi \to 2\pi$'' show the free SIMP freeze-out without and with the WZW interactions, while the lower dashed gray line corresponds to the observed DM abundance. Vertical lines give the corresponding values of $x_1$ (decoupling point of mass depletion processes) and $x_2$ (decoupling point of $3\pi \leftrightarrow \pi X$) with matching colors. In all panels $x_1 \le x_2$ holds.
}
\label{fig:varysolutions}
\end{figure}

The analytical solutions provided in the main text, Eq.~\eqref{eq:evenBoltzSol} and~\eqref{eq:oddBoltzSol}, to  even- and odd-numbered cases, indicate the parametric dependencies of model parameters. In Fig.~\ref{fig:varysolutions},  we test numerically how the evolution of the total DM abundance changes with varying parameters and  compare with our analytical solutions. 
Concretely, the parameters adopted for our benchmark model in the main text yield
\begin{align}
    \sigma^{\rm LO}_{\rm el}/m_\pi & \simeq   {0.2}\,\text{bn}/\text{GeV} \simeq {0.1}\,\text{cm}^2/\text{g}\,,\\
    \langle \sigma_{2X \to 2\pi} v^2 \rangle/m_\pi  &  \simeq {6.5}\,\text{bn}/\text{GeV}\,, ~~\langle \sigma_{\pi X_p  \to 2\pi} v \rangle/m_\pi  \simeq {0.03 x^{-1}}\,\text{bn}/\text{GeV}\,.
\end{align}
For the even-numbered case, where the final chemical decoupling of bound state formation, $3\pi \to \pi X$, happens at $x_2 \simeq 115$, our analytical solution, Eq.~\eqref{eq:evenBoltzSol},  yields  $Y(x_2) \simeq 0.35Y|_{\rm num}$, while its approximation,  Eq.~\eqref{eq:BoltzSolApp}, yields $Y(x_2) \simeq 0.28Y|_{\rm num}$; here $Y|_{\rm num}$ denotes the final yield of the full numerical solution. We see that the simple analytical expression works to within a factor of 2-3.
More parameter choices for the even-numbered case are shown in  the left panel of Fig.~\ref{fig:varysolutions}, where we change $R(0)/ m_\pi^{3/2} = 0.35$ to $ 0.7$ and $ 1.4$, of which the last yields approximately the observed DM relic abundance.
For the last parameter choice, one obtains  $x_2 \simeq  111$, so Eq.~\eqref{eq:evenBoltzSol}  yields $Y(x_2) \simeq 0.39Y|_{\rm num}$, while the approximation yields $Y(x_2) \simeq 0.27Y|_{\rm num}$. Interestingly, the relative differences between the analytical and numerical results are similar in both sets of parameters. 

To study the parameter-dependence quantitatively,  increasing  $R(0)$ by a factor of $2$ enhances the relevant process $\langle \sigma_{2X \to 2\pi} v^2 \rangle$ by a factor of $2^4 = 16$. Our analytical expression suggests that this change reduces the final abundance by $16^{1/3} \simeq 2.5$ if the freeze-out point~$x_2$ remains. Indeed, our full numerical solution shows an decrease of the final abundance by about a factor of $2.2$ in the left panel of Fig.~\ref{fig:varysolutions}.  Finally, we have also numerically calculated how the relic abundance changes by  doubling  the binding energy to $\kappa =0.2$. Although not demonstrated explicitly with figures here, our numerical results show that this reduces $x_2$ by roughly a factor of~$2$. The analytical solution then suggests a suppression factor of $2^{-4/3} \simeq 0.40$ of the final DM abundance. Indeed, the full numerical results  yield a very similar factor~$0.39$.  
In summary, our analytical expression of the even-number case estimates the final abundance within a factor of $2-3$ and shows excellent scaling behavior with respect to a variation of model parameters.

\subsection{Odd-numbered case: analytical and numerical solutions}

\begin{figure}%
\includegraphics[width=0.32\textwidth]{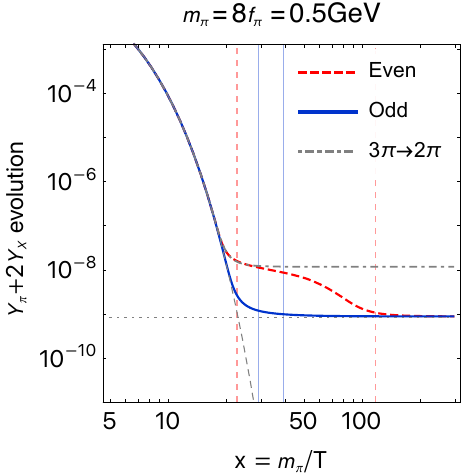}
\hfill
\includegraphics[width=0.32\textwidth]{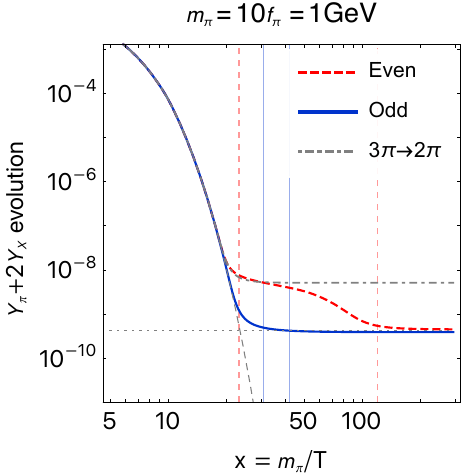}
\hfill
\includegraphics[width=0.32\textwidth]{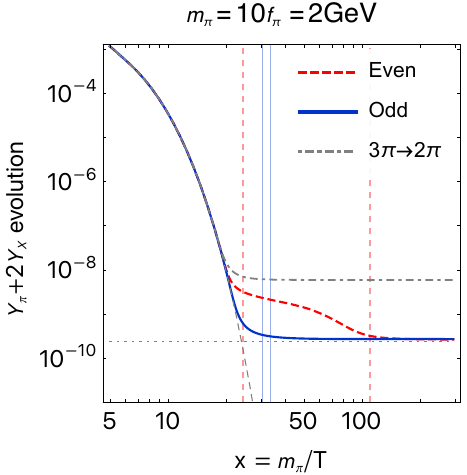}
\caption{
Evolution of DM abundance, $Y_{\pi} + 2Y_{X}\simeq Y_{\pi} $ for various DM masses with parameters chosen to achieve the observed relic abundance. \emph{Left}: $m_\pi = 8f_\pi = 0.5$\,GeV, for even case $R(0)/ m_\pi^{3/2}= 1.1$, and for odd case $(R(0)/ m_\pi^{3/2}, R'(0)/ m_\pi^{5/2}) = (0.2,\,0.02)$.  \emph{Middle}: $m_\pi = 10f_\pi = 1$\,GeV,  even case $R(0)/ m_\pi^{3/2}= 1.3$, and odd case $(R(0)/ m_\pi^{3/2}, R'(0)/ m_\pi^{5/2}) = (0.2,\,0.02)$.  \emph{Right}: $m_\pi = 10f_\pi = 2$\,GeV,  even case $R(0)/ m_\pi^{3/2}= 2.6$, and odd case $(R(0)/ m_\pi^{3/2}, R'(0)/ m_\pi^{5/2}) = (0.2,\,0.06)$.   We fix the binding energies to $\kappa_S =4\kappa_P = 0.1$. The leading-order DM self-scattering cross section per mass  $\sigma_{\rm el}^{\rm LO}/m_\pi = 0.08$,~$0.03$,~$0.003$~bn/GeV, respectively, from left to right. Other curves are the same as in Fig.~\ref{fig:varysolutions}. }
\label{fig:LargeMass}
\end{figure}

Now we turn to the odd-numbered case. Here, the DM abundance essentially already freezes out at $x_1$, since $\pi X_p \to \pi\pi$ decouples with exponential temperature dependence. The relic abundance can be estimated from $n_{X_p} \langle\sigma_{\pi X \rightarrow 2\pi} v\rangle \simeq H(x_1)$. When $3\pi \leftrightarrow \pi X$ is strong enough at $x_1$, there exists the detailed balance condition $Y_X  = {Y_\pi^2 Y_X^{\rm eq}}/{(Y_\pi^{\rm eq})^2}$. Using this equality at $x_1$, we obtain 
\begin{equation}
\label{eq:oddBoltz}
    Y_\pi(x_1) \simeq  \left(\frac{91125 N_\pi^4 e^{-2\kappa_P x_1}  x_1^5}{2048 \pi^9 g_*^3 N_{X_p}^2   M_P^2 m_\pi^2 \langle\sigma_{\pi X_P \rightarrow 2\pi} v\rangle^2  } \right)^{1/4} \, ,
\end{equation}
which, upon evaluation, yields Eq.~\eqref{eq:oddBoltzSol} of the main text. Comparing this analytical solution to the numerical results is straightforward, as we show explicitly the vertical line where $n_{X_p} \langle\sigma_{\pi X_P \rightarrow 2\pi} v\rangle \simeq H(x_1)$ is reached in the middle and right panels of Fig.~\ref{fig:varysolutions}. We find that $Y_{\pi} (x_1)$ is already  within a factor of two of its asymptotic value at $x \gg x_1$.
Another way to estimate the DM relic abundance in  the odd-numbered case is to integrate the both sides of the approximate differential equation
\begin{equation}
\frac{d  Y_{\pi}}{d x} \simeq    - \frac{s}{H x} 
\langle\sigma_{\pi X_P \rightarrow 2\pi} v\rangle \frac{Y^{\rm eq}_X}{(Y^{\rm eq}_\pi)^2 } Y_{\pi}^3  \,,
\end{equation}
which actually leads to an analytical expression very similar to Eq.~\eqref{eq:oddBoltz} for $\kappa_P x_1 \lesssim 1$.

Note that there also exists another distinct possibility for very large values of $R'(0)$, where the detailed balance at $x = x_1$ is rather established between the formation of~$X$, via $3\pi \to \pi X $, and the depletion of~$X$, via $\pi X_p \to 2\pi $. In this case, we, in turn, have
\begin{equation}
 Y_{X_{(P)}}^{\rm eq} \left(\frac{Y_\pi}{Y_\pi^{\rm eq}} \right)^2  > Y_{X_{(P)}} >    Y_{X_{(P)}}^{\rm eq} \left(\frac{Y_\pi}{Y_\pi^{\rm eq}}\right) \,. 
\end{equation}
This happens for the parameter choice of $2R(0)/m_\pi^{3/2} = R'(0)/m_\pi^{5/2} = 0.06$ (red dotted curve) in the  middle panel of Fig.~\ref{fig:varysolutions}, leading to  $x_1 = x_2$. Nevertheless, in reality we expect   $R(0)/m_\pi^{3/2} \gg R'(0)/m_\pi^{5/2}$ for bound states to typically hold. We leave a quantitative study of this possibility for future work. %

\begin{figure}
\includegraphics[width=.49\columnwidth]{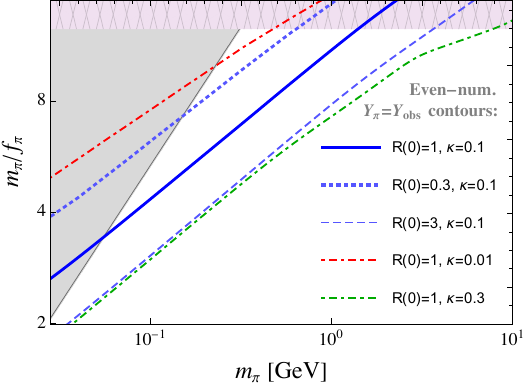}
\includegraphics[width=.49\columnwidth]{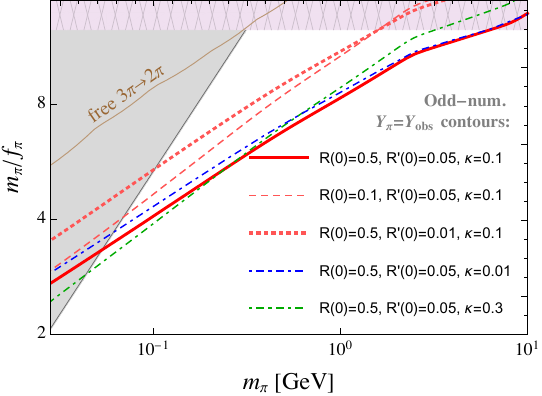}
\caption{
Parameter scan of even-numbered (\emph{left}) and odd-numbered (\emph{right}) cases, for various values of $R(0)$ in units of $m_\pi^{3/2}$, $R'(0)$ in units of $m_\pi^{5/2}$, and $\kappa$ that yield the observed DM abundance. The gray-shaded region shows DM self-scattering exclusion $\sigma^{LO}_{\rm el} \ge 2$\,bn/GeV; in the top hatched region, $m_\pi/f_\pi>4\pi$. The free  $3\pi \to 2\pi$ freeze-out via the WZW interactions for $Y_{\rm obs}$,  the brown line in the right panel, only appears at the top-left corner, while the line corresponding for $4\pi \to 2\pi$ freeze-out does not appear in the parameter region,  requiring $m_\pi/f_\pi > 4\pi$ for $m_\pi < 1$\,MeV.  
}
\label{fig:oddscan}
\end{figure}

\subsection{Prospective DM mass range} 

Finally, we explore the prospective higher mass DM range. The simple model considered here is very restrictive as it only has a few principal parameters such as $m_\pi$ and $m_\pi/f_\pi$. Larger masses $m_\pi$ require larger interactions, i.e., larger ratios of $m_\pi/f_\pi$ to achieve the observed relic abundance. One enters a regime where next-to-leading order and unitarizing corrections become important. Here we simply explore the general trends based on our leading-order formulation. In Fig.~\ref{fig:LargeMass} we show examples for $m_\pi = 0.5\,\GeV$ (left panel), $m_\pi = 1\,\GeV$ (middle panel) and $m_\pi = 2\,\GeV$ (right panel). Note that in all examples, the  perturbative limit, $m_\pi/f_\pi < 4\pi$, and the condition for the  longevity of bound states, $|\Psi(0)| = R(0)/\sqrt{4\pi} <m_\pi^{3/2} $, are satisfied. 

Full parameter scans are given in the main text for the even-numbered cases, as well as in Fig.~\ref{fig:oddscan} for both cases. In the odd-numbered case (right panel) the dependence of the final abundance on $\kappa$ is both mild and subtle. Larger $\kappa$ leads to a weaker enhancement $(m_\pi/E_B)^{3/2}$ of the resonant process $3\pi \to \pi X$, but to  a larger thermal population of  bound states because of the lighter $X$-mass. The latter helps to maintain $\pi X_P \to \pi\pi$ longer. In contrast, in the even-numbered case (left panel),  the bottleneck for freeze-out is typically the $X X \to \pi\pi$ process, so increasing $\kappa$, and thus increasing $Y_X^{\rm eq}$, can greatly reduce the relic abundance in the final stages of freeze-out. This is shown by the green dash-dotted line.

\end{document}